\documentclass[a4paper,twocolumn,10pt]{article}
\usepackage[left=1.5cm,top=3cm,right=1.5cm,bottom=3cm]{geometry} 
\usepackage{epsfig}
\usepackage{multicol}
\usepackage[latin1]{inputenc}
\usepackage{graphicx}
\usepackage{natbib}
\usepackage{caption}
\usepackage{wasysym}
\usepackage{tikz}
\usepackage{hyperref}
\usepackage{upgreek}
\usepackage{cite}
\usepackage{transparent}
\usepackage{eso-pic}

\hypersetup{
	colorlinks=true,
	linkcolor=red, 
	urlcolor=orange,
	citecolor=blue,
	bookmarks=true,
}

\begin{document}

\title{Evolutionary process of the interacting binary V495 Centauri}
\author{
Rosales, J.~A.$^{1}$, Mennickent, R.~E.$^{1}$, Schleicher, D.~R.~G.$^{1}$, Senhadji, A.~A.$^{2}$\\ \\
\small{$^{1}$ \emph{Departamento de Astronomía, Universidad de Concepción, Casilla 160-C, Concepción, Chile.}}\\
\small{$^{2}$ \emph{Physics and Astronomy Department, Bishop's University, Sherbrooke, Canadá.}}\\
}
\date{}
\vskip 2cm

\twocolumn[
\begin{@twocolumnfalse}
	\maketitle
\begin{abstract}

We present a simple model for the double periodic variable (DPV) V495 Cen, which evolves as a binary system of intermediate mass, where the gainer cannot accrete at high rate, limited by the Eddington accretion rate, leading to the formation of an accretion disc. The theoretical model begins at the zero age main sequence considering the rotation for both stars. For this purpose we used the stellar evolution code {\textsc{mesa}}, developed to calculate the evolution of stars in a wide range of parameters. We started the model adjusting fundamental parameters published for this system through a chi-square optimization algorithm, and adopting an initial orbital period of 3.9 days and initial masses for the primary component $M_{i,d} = 3.40$ $M_{\odot}$ and $M_{i,g}= 3.18$ $M_{\odot}$ for the gainer, with a metallicity associated to this type of DPV of $Z = 0.02$. The method converged successfully for eight free degrees and 5\% of confidence with a chi-square of $\Delta \chi^{2}_{0.95,8}= 0.212$. We describe each evolutionary stage of both components until that the donor reaches 20\% core helium depletion as stop criterion. We offer a complementary analysis for understanding the mechanism of the magnetic dynamos inside the donor star using the Tayler-Spruit formalism. Currently, the theoretical model is consistent with the fundamental parameters published for V495 Cen and we discuss how our predictions can help to develop efficients theoretical models for DPV stars.\\

{\bf Keywords:} Binaries: general- stars: evolution\\ \\
	\end{abstract}
\end{@twocolumnfalse}
]
\vskip 0.5cm
{\bf 1 INTRODUCTION}\\

The evolution of Double Periodic Variables (DPV) has been little studied to date. Our understanding of the internal structure of each component is based on spectroscopy and photometric studies of the last years. These system are binary stars of intermediate mass that show two photometric variations, where the long period is on average about 33 times longer than the orbital period \citep{2003A&A...399L..47M,2017SerAJ.194....1M}. A property of these systems is the constancy of their orbital period, which is not expected in classical Algols undergoing Roche Lobe Over Flow (RLOF) mass transfer \citep{2013MNRAS.428.1594G}, mainly because intermediate mass binaries evolve in a conservative way, but the mass will be blown away from the system during the short era of rapid mass-transfer soon after the onset of RLOF \citep{2008A&A...487.1129V}. The increase of mass loss from a binary system is related to the initial orbital period and the initial mass of the donor. I.e., if we compare two binary systems of equal total masses, then the system that present the largest initial orbital period will be the one who loses more mass. The same occurs at the case of the initial mass of the donor \citep{2011A&A...528A..16V}.

The Algol system already presented a cycle of 32 year, reported by \citet{1980A&A....89..100S}, which was not linked to the orbit of the system. Years later, \citet{1987ApJ...322L..99A} proposed a mechanism to produce the observed long-term period variations as result of magnetic activity cycles, produced on the external convection zone of one star of the system, responsible for the changes in the orbital period on a time scale of order 10 years. Then \citet{1997MNRAS.286..209S}, examined the possibility of tracking the existence of dynamos in Algol type binary stars that show mass loss. Later \citet{2004MNRAS.352..416M} described episodes of magnetized mass transfer in the cataclysmic variable AE Aquarii, subsequently \citet{2010Natur.463..207P} reported magnetic activity in a close binary Algol, where the most evolved component is a bright radio active KIV star while its companion  is a star of the main sequence of spectral type B8. They suggested that both stars are aligned by a persistent asymmetric magnetic field between both stars. Currently, the second period is linked to mechanisms based on magnetic dynamo cycles of the donor star is proposed by \citet{2017A&A...602A.109S} and variations of the wind generated by impact of a gas stream onto the accretion disc \citep{2012IAUS..282..317M, 2016MNRAS.461.1674M}. The DPV V495 Centauri is currently studied through photometric and spectroscopic analysis by \citet{2018MNRAS.476.3039R}. The system includes an evolved star with characteristics of an F-type star, transferring mass onto an early B-type dwarf, hidden by an optically and geometrically thick disc of radial extension $R_{d}=40.2 \pm 1.3$ $R_{\odot}$. Some one the characteristics of this system are that it has an extremely long period of 1283 days, shows a persistent V $<$ R asymmetry of the H$\alpha$ emission line and shows also the presence of a possible wind formed in the hotspot region. The fundamental parameters proposed for this system include an accretion disc around the gainer star, which is in agreement with the spectral energy distribution that shows infrared excess indicating the presence of circumstellar material. We offer a complementary analysis of the system V495 Cen to understand the evolution of stars in a binary system as a DPV, applying the modern stellar evolution code {\textsc{mesa}}\footnote{http://mesa.sourceforge.net/} \citep{1971MNRAS.151..351E,2004PASP..116..699P,2011ApJS..192....3P,2013ApJS..208....4P} developed to calculate evolution of stars in a wide range of parameters, with independent modules for experiments in stellar astrophysics of close binary systems \citep{2015ApJS..220...15P,2018ApJS..234...34P}.\\

\vskip 0.5cm
{\bf 2 ON THE MODEL}\\

When the primary star expands beyond a critical volume called Roche lobe and increases its radius, it begins eventually to interact with the companion. The module for experiments with a close binary star of {\textsc{mesa}} is developed to find an analytic approximation of the Roche lobe around of the primary star, according to:\\

\begin{equation}
R_{L}= a \frac{0.49 q^{2/3}}{0.6 q^{2/3}+ln(1+q^{1/3})},
\label{eq: eq. 1}
\end{equation}
\\

\noindent
(\citet{2006epbm.book.....E}, eq. 3.5) where $R_{L}$ is the effective radius for the critical equipotential of the Roche lobe and $q=m_{2}/m_{1}$ the mass ratio. Normally, in the binary system we distinguish three cases of mass transfer based on the evolutionary stage of the donor star, when it fills the Roche lobe, defined by \citet{1967ZA.....65..251K} and \citet{1970A&A.....7..150L}:\\

(i) Case A : The donor star evolves first filling the Roche lobe during the core hydrogen burning. Here the evolution is slow (nuclear timescale), and during its process of mass transfer the donor star evolves in a stable way along of the main sequence.\\

(ii) Case B : The donor has filled the Roche lobe and leaved the main sequence. This process occurs before core helium ignition, the mass transfer is driven by the fast expansion of the donor in a thermal timescale since the donor is crossing the Hertzsprung-gap while is burning hydrogen in a shell.\\

(iii) Case C : The donor star has filled the Roche lobe after the core helium ignition, both stars interact before the donor ends its life. In high mass stars the mass transfer occurs in a dynamical timescale and it is often unstable.\\ 

\begin{figure}
	\begin{center}
	\includegraphics[width=8 cm,angle=0]{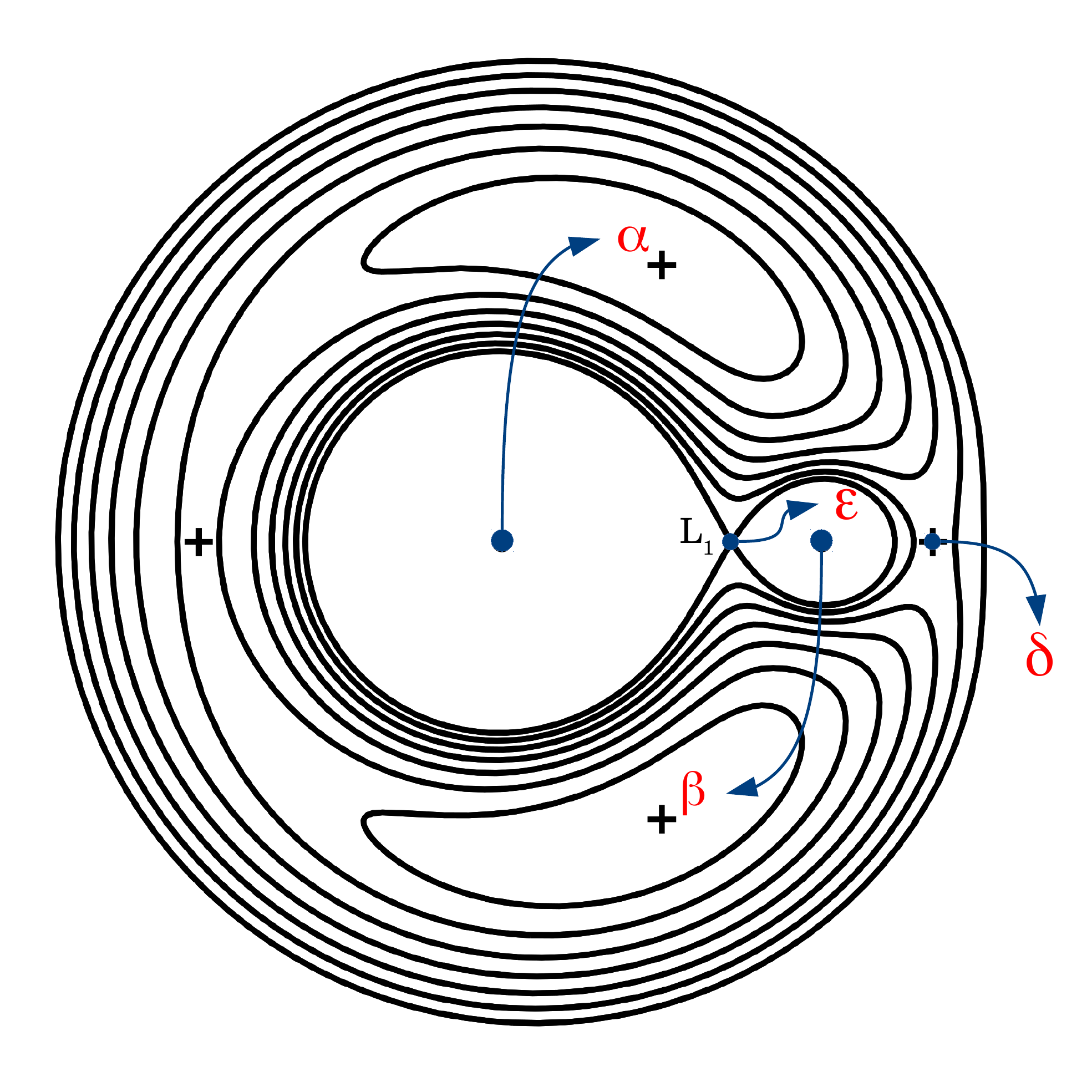}
	\end{center}
	\caption{Mass loss/transfer fraction. $\alpha$ correspond to the fraction of mass lost from the vicinity of the donor as fast wind, $\beta$ mass lost from the vicinity of the accretor, $\delta$ is the mass lost from the circumbinary coplanar toroid  and $\epsilon$ is the fraction of accreted mass.}
	\label{Fig:Fig.1}
\end{figure}

\vskip 0.5cm
{\bf 2.1 MASS TRANSFER}\\

Particularly, the mass transfer produces important effects in the stars. This process can be accompanied by stellar winds from the mass-losing star, or ejection of matter from the accretor \citep{1997A&A...327..620S}. In our calculation, we take into account the following mass loss/transfer fractions (see Fig.\ref{Fig:Fig.1}); $\alpha$: the Jeans's mode is a fast mode, associated to a spherically symmetric outflow from the donor star in form of a fast wind; $\beta$: at the isotropic re-emission, the flow is transported from the donor star to the vicinity of the accretor, where it is ejected as a fast isotropic wind; $\delta$: this mode is considered an intermediate mode or mass loss to a ring. The ejection of the mass has the characteristic of the angular momentum of a circumbinary ring; $\epsilon$: is the fraction of the mass accreted by the secondary. The donor star has filled the Roche lobe, here the matter is lost from vicinity of the donor, through the inner Lagrange point, to the vicinity of the accretor, which arrives with high angular momentum. However, the $\delta$ mode was not considered because a circumbinary coplanar toroid, is not observed in the DPV V495 Cen.

When the donor expands, the flow from the outermost layers accelerates and converges towards the Lagrangian point $L_{1}$ ( Fig. 1), it reaches sound velocities near the point $L_{1}$, and fall onto the gainer with a mass transfer rate given by \citet{1990A&A...236..385K}:\\

\begin{equation}
	\dot{M}= \int \rho_{L} \nu_{s} dA,
\label{eq: eq. 2}	
\end{equation}
\\

\noindent
where $\rho_{L}$ and $\nu_{s}$ are the density and the sound speed of the flow at the $L_{1}$ point, respectively. But, the mass transfer from the primary star (donor) to the companion is limited by the Eddington accretion rate. This accretion represents the rate at which the radiation produced by the heated infalling material is strong enough to prevent the accretion:\\

\begin{equation}
	\dot{M}_{Edd}= \frac{4\pi c R}{\kappa},
\label{eq: eq. 3}	
\end{equation}
\\

\noindent
(\citet{1983ApJ...265..393B}, eq. 3),  where $R$ is the radius of the accreting star and $\kappa$ is the opacity. {\textsc{mesa}} assumes the mass losses in a stellar wind has a specific orbital angular momentum. Therefore for an inefficient mass transfer, where fixed fractions of the transferred mass are lost either as a fast isotropic wind from each star or a circumbinary toroid are compatible. Also {\textsc{mesa}} always ensures that the mass loss/transfer are always consistent with each other, as both are calculated self-consistency, i.e. this provides both explicit and implicit methods to compute mass transfer rates. An explicit computation sets the value of $\dot{M}_{RLOF}$ at the start of a step, while an implicit one begins with a guess for $\dot{M}_{RLOF}$ and iterates until the required tolerance is reached. The composition of accreted material is set to that of the donor surface, and the specific entropy of accreted material is the same as the surface of the accretor \citep{2015ApJS..220...15P}.\\

\vskip 0.5cm
{\bf 2.2 MASS LOSS}\\

After adjusting the mass transfer of the binary system, we have implemented the mass loss due to winds. We used the model of \citet{1975MSRSL...8..369R} for the mass loss in red giants,  while for AGB stars towards the stage of the white dwarf the mass loss will be considered based on dynamical calculation of the atmospheres of Mira-like stars proposed by \citet{1995A&A...297..727B}. In addition, for a range of stars within the H-R diagram, the mass loss is represented by a function depending only on T$_{eff}$ and L of the stars proposed by \citet{1988A&AS...72..259D}, but when we deal with mass loss for massive stars is recommended to use the models of \citet{1990A&A...231..134N}. However, in another cases for stars like Wolf-Rayet  the model of \citet{2000A&A...360..227N} is implemented. Unlike  the previous case of mass loss in O and B stars, where it is related with their metallicities \citep{2001A&A...369..574V},  while the massive stars can be worked with the model of  \citet{2009A&A...497..255G}. 

In other cases such as supersonic mass loss one could consider the formalism by \citet{1995ApJ...445..789P}, and when we deal with super Eddington mass loss we should use the formalism by \citet{1986ApJ...302..519P}. Finally, to ensure the correct representation of the mass loss process in intermediate mass stars through the He-core burning phase, we have to consider time steps small enough to allow the convergence of the results.\\

\vskip 0.5cm
{\bf 2.3 INTERNAL STRUCTURE AND ROTATION}\\

The relevance of the rotation in the stars is that the centrifugal forces interacting with the matter drive deviations from spherical symmetry. It is known that large mean rotational velocities are common among the early-type stars and that these velocities decline steeply in the F-star region, from 150 km s$^{-1}$ to less than 10 km s$^{-1}$ \citep{2000stro.book.....T}. The instabilities discussed in this section are a list of all rotationally induced instabilities, relevant for stellar evolution of intermediate mass within our model, which are included in {\textsc{mesa}} code. However, they are well studied by \citet{2000ApJ...528..368H,2005ApJ...626..350H} and are in general relevant for the present case. 

The dynamical shear instability (mixing process) \citep{1992A&A...265..115Z} is an instability that occurs when the energy that is gained from the shear flow becomes comparable to the work that has to be done against the gravitational potential for the adiabatic turnover of a mass element, i.e., this means that this instability is stabilized by density gradients. Another type of instability related to adiabatic process is the called Solberg-Hoiland (SB) \citep{1946ApNr....4....1W} instability, which arises if an adiabatically displaced mass element experiences a net force that has components in the direction of the displacement. It is a convective instability type favored by stellar rotation, just if the outward decrease in angular velocity is moderate according to the Rayleigh criterion:\\

\begin{equation}
	\frac{dj}{d\overline{\omega}}= \overline{\omega}\left(\overline{\omega}\frac{d\Omega}{d\overline{\omega}}+2\Omega\right)>0 \hspace{0.5cm} \textrm{or} \hspace{0.5cm} \frac{dln{\Omega}}{dln\overline{\omega}}>-2
\end{equation}

\noindent
where $j= \overline{\omega}^{2}\Omega$ is the angular momentum, $\Omega$ angular velocity and $\overline{\omega}$ is the distance from rotational axis until the element fluid \citep{2009pfer.book.....M}. Despite that we have not used this instability for the internal structure of our stars we consider that is necessary to be mentioned due to the relationship that it has with the previous instability and the accretion discs. 

Another interesting instability that can occur in radiative stellar zones due to the presence of inverse gradients of the mean molecular weight $\mu$, is the thermohaline convection. Here the mean molecular weight gradient  $\nabla_{\mu}=\textrm{dln}{\mu}/\textrm{dln}{P}$ plays the role of the gradient of the average mass per particle relative to the mass of hydrogen, while the difference between the adiabatic term and the local (radiative) gradient $\nabla_{ad}-\nabla$ plays the role of the temperature \citep{2008IAUS..252...97V}, i.e., this mechanism governs the photospheric composition of bright giants of low mass, and is a type of double diffuse instability which is observed as elongated fingers. This instability is induced by the molecular weight inversion created by the $^{3}$He + $^{3}$He $\longrightarrow$ $^{4}$He + 2$^{1}$H reaction in the outer zone of the hydrogen combustion layer, converting two particles to three as predicted by \citet{1972ApJ...172..165U}. It is expected that it will stabilize after the first dredging, when the star reaches the RGB brightness. 

When the radiative gradient $\nabla_{rad}$ is intermediate between the stability predicted by Ledoux criterion and the instability predicted by the Schwarzschild criterion it is called semiconvection:\\

\begin{equation}
	\nabla_{int}<\nabla<\nabla_{int}+ \left(\frac{\varphi}{\delta}\right)\nabla_{\mu},
	\label{eq: eq. 4}
\end{equation}

\ \\
\noindent
with $\nabla_{int} \approx \nabla_{ad}$ and $\nabla \approx \nabla_{rad}$ \citep{2009pfer.book.....M}. The semi convection \citep{1966PASJ...18..374K} is a type of instability that can occur in non rotating stars. It is a secular shear instability, i.e., this type of  instability appears in regions where an unstable temperature gradient is stabilized against convection by a sufficiently large gradient of mean molecular weight. However, we known that our objects are rotating stars and, therefore, we have considered the Eddington-Sweet circulation \citep{1959ZA.....48..140B}, which considers that a rotating star cannot be in hydrostatic and radiative thermal equilibrium at the same time. This is because surfaces of constant temperature and constant pressure do not coincide. Consequently, large-scale circulations develop. Since inhomogeneities on isobars are quickly smoothed out by the horizontal turbulence only the perpendicular component of the circulation velocity is considered here, and the process is approximated by diffusion along the radial coordinate.

Inasmuch as the temperature plays a fundamental role in the stars, the Goldreich-Schubert-Fricke (GSF) instability \citep{1967ApJ...150..571G,1968ZA.....68..317F} is essential to to study the instability in rotating stars whose profiles show gradients along radial and vertical coordinates. It is composed by two criteria, the first is related to the stabilization by the temperature gradient, which is removed due to thermal conduction. The second criterion is analogue to Taylor-Proudman \citep{1974IAUS...66...20K,1978trs..book.....T} for slowly rotating incompressible fluids. But it is thought that  GSF is probably far less efficient in the transport of angular momentum than it is often assumed \citep{2016MNRAS.460..338C}.  Therefore the condition for this instability in the limit of no viscosity and no stratification is:\\

\begin{equation}
	\left |\frac{\partial(rv_{o})}{\partial r}\right| > \left| \frac{l_{r}}{l_{z}} \frac{\partial(rv_{0})}{\partial z} \right|,
	\label{eq: eq. 5}
\end{equation}

\ \\
\noindent
where $l_{r}$ and $l_{z}$ are the radial and vertical length scales of the system \citep{2016mfdp.book.....R} and $r$ represents a radial boundary. The criterion says, if the vertical gradient of the swirling flow is weak, then this instability is expected only for short radial wavelength disturbances. 

One of the more effective ways of transport of angular momentum and magnetic field amplification is the Spruit-Tayler dynamo proposed by \citet{2002A&A...381..923S}, which provides estimates for the magnetic field components B$_{r}$ and B$_{\phi}$ based on dynamo processes that take into account the effect of stable stratifications. This model assumes that the rotation rate is a function of the radial coordinate and the initial magnetic field must be sufficiently weak to neglect the initial magnetic forces and allow the radial component B$_{r}$ to be wounded by the differential rotation, later of some turns this would be predominately azimuthal i.e., $B_{\phi}$ $>$ B$_{r}$ and increases linearly until be unstable. Therefore in this case the differential rotation is sufficiently strong to maintain a dynamo process. Then, the azimuthal field produced must be large enough to generate the Tayler instability. Considering an instability without thermal diffusion overestimates the intensity of the field required for this, if the stratification is due to the thermal gradient. When the stabilizing stratification is due to the entropy gradient, the heat is transported by the photons. While the viscocity and magnetic fields are diffused by Coulomb interactions, known as the Prandtl number, here the viscocity is of the same order as the magnetic diffusivity, and at the same time it is the mechanism that controls the magnetic diffusion. Therefore the stabilizing effect of the stratification due to the entropy gradient is reduced by thermal diffusion, and the instabilities may appear more easily. Independent of the intensity of the azimuth field, the condition derived by Spruit is:\\

\begin{equation}
	\frac{\omega_{A}}{\Omega}> \left(\frac{N}{\Omega}\right)^{1/2}\left(\frac{\kappa}{r^{2}\Omega}\right)^{1/4}\left(\frac{\eta}{\kappa}\right)^{1/2},
	\label{eq: eq. 6}
\end{equation}

\ \\
\noindent
where the conditions to validate the previous equation are $\omega_{a} << N << \Omega$ and $\eta<<\kappa$. In the above equation $\eta$ is the magnetic diffusivity (cm$^{2}$s$^{-1}$), $N$ is the buoyancy frequency, $\Omega$ the rotation rate, $\kappa$ is the thermal diffusivity and r is the radial length scale \citep{2002A&A...381..923S}. Also the minimum shear rate $q$ required to produce the critical magnetic field strengths for Taylor instability and drive convective motions are:\\

\begin{eqnarray}
	q_{0} & = & \left(\frac{N}{\Omega}\right)^{7/4} \left(\frac{\eta}{r^{2}N}\right)^{1/4}\\
	q_{1} & = & \left(\frac{N}{\Omega}\right)^{7/4} \left(\frac{\eta}{r^{2}N}\right)^{1/4}\left(\frac{\eta}{\kappa}\right)^{3/4},
	\label{eq: eq. 7}
	\label{eq: eq. 8}
\end{eqnarray}

\ \\ 
\noindent
wherein $q$ is the dimensionless differential rotation rate when the thermal diffusion can be neglected $q_{0}$, and when the thermal diffusion is considered $q_{1}$, i.e., when the effects of stratification is dominated by the composition gradient. The overshooting process depends on the adjacent convective layers, where the acceleration of the cells is zero or null, i.e., a neutral stability corresponds when $\nabla_{rad}= \nabla_{ad}$, but strictly speaking the temperature gradient must be slightly sub-adiabatic, otherwise the convective elements never show deceleration \citep{2012sse..book.....K}, this process is also known by other authors as convective penetration. The difference  that exists between kinematic and dynamic edges is the average overshooting distance $d_{over} = |{r_{v} -r \Delta t}|$ to which the convective mixture extends, beyond the formal limit defined by Schwarschild or Ledoux \citep{2009pfer.book.....M}. When the rotation is present in a convective core the SH criterion and the semi convection could influence the overshooting process, in turn the overshooting will determine the amount of nuclear fuel for the star and thus directly affecting the age of the star \citep{2007ApJ...667..448M}. Another effect that can extend the convective core is the rotational mixing, which conducts more hydrogen in the center, which can change the global properties of a solar-type star with a significant increase of the effective temperature, producing a change in the evolutionary track of the star due  to the rotational mixing counteracts the effects of atomic diffusion \citep{2010A&A...519A.116E}. In addition the mixing effects grow with age, because the mixture has more time to influence the distribution of elements \citep{2009pfer.book.....M}.

We want emphasize that the Tayler-Spruit dynamo is certainly a fundamental ingredient among the processes considered here, as it is the only process that can generate a magnetic field, and which may also drive convection due to magnetic instabilities. As we will show in the next chapters, thermohaline convection is also quite important during the later phases of binary evolution, though more for the gainer rather than for the donor. The models employed here are based on approximate descriptions, as the processes themselves are highly non-linear. While in principle a 3D treatment would be required, the latter is not feasible if the system if followed over a significant portion of the lifetime of the star. The results therefore have to be taken \textquotedblleft cum grano salis \textquotedblright, as a fully consistent treatment is not possible at present.

\ \\
\vskip 0.5cm
{\bf 2.4 THE FITTING PROCEDURE}\\

We present a {\textsc{mesa}} model which evolves a binary system of intermediate mass, and we have specified to {\textsc{mesa}} how should treat the mass transfer from the donor star to the companion, which cannot accrete at a high rate, limited by the Eddington accretion rate, i.e., simulating the formation of an accretion disc on the gainer star. The theoretical model begins at the zero-age main sequence with rotation for both stars, but we have constrained the rotation of the primary star (donor) to 26 km s$^{-1}$ as initial conditions to never reach high rotation rates during their evolution. Also we consider a binary interaction with practically almost conservative processes. We initiated the model using parameters previously derived for the DPV V495 Cen (\citet{2018MNRAS.476.3039R}, Table 2), adopting an initial orbital period 3.6 to 4.6 d with steps of 0.1 d, an initial mass for the primary component (donor) of 3.0 to 5.1 M$_{\odot}$ with step of 0.01 M$_{\odot}$ and for the companion (gainer) of 1.9 to 4.0 M$_{\odot}$ with steps of 0.01 M$_{\odot}$. The metallicity associated to the DPV is $Z=0.02$. Also, both stars include differential rotational and the controls to generate a dynamo model considering a Solberg-Hoiland (SH) instability, the diffusion coefficient for angular momentum, secular shear instability (SSI), Eddington-Sweet (ES) circulation, and the Spruit-Tayler (ST) dynamo. The surface rotation for the primary component is 26 km s$^{-1}$, while that for the companion the surface rotation is a free parameter that can be estimated by {\textsc{mesa}} assuming synchronous rotation. The principal stopping criterion for the evolution is when the primary star reaches core helium depletion ($X_{He_{4_{c}}} < 0.2$). We considered  a semiconvection according to the Ledoux criterion \citep{2009CoAst.158..277L} where the stability occurs if the radiative gradient  is less than adiabatic gradient $\nabla (r) < \nabla (L)$. The convective overshooting and a wind mass loss are considered according to \citet{1988A&AS...72..259D} for a cool wind described by \citet{1995A&A...297..727B}.

\begin{figure}[h!]
	\begin{center}
		\includegraphics[width=8 cm,angle=0]{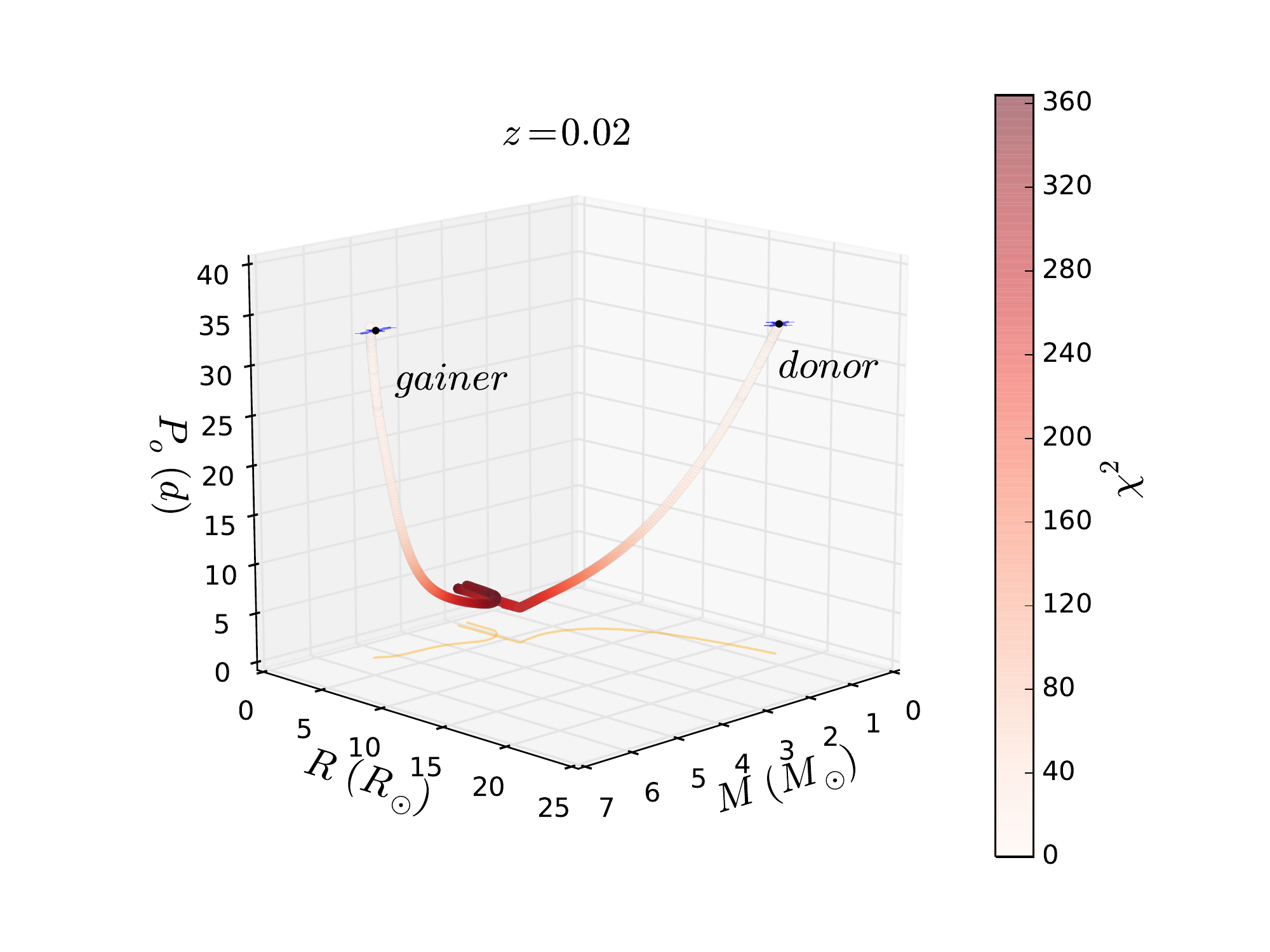}
	\end{center}
	\footnotesize
	{\bf Figure 2.} Best model for V495 Cen of 8 freedom degree with all optimized parameters values for both stars, with a $\Delta \chi^{2}_{0.95,8}= 0.212$.
	\normalsize 
\end{figure}

\begin{figure}
	\begin{center}
		\includegraphics[width=7.2 cm,angle=0]{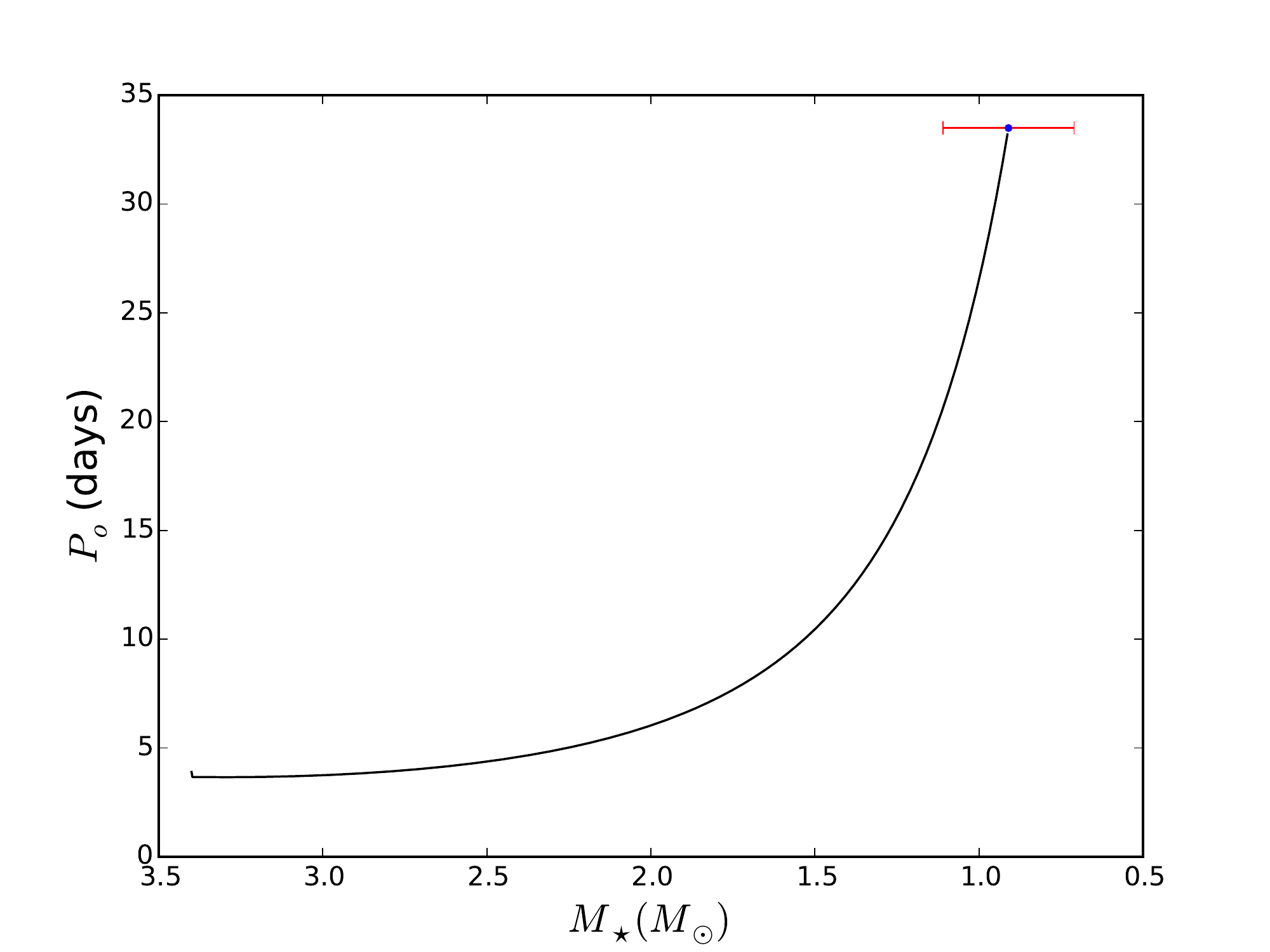}
	\end{center}
	\footnotesize
	{\bf Figure 3.} Evolution of the orbital period as a function of the mass of the primary star (donor) of V495 Cen. 
	\normalsize 
\end{figure}

Our model is aimed to find the best fit, which was based on a chi-square optimization algorithm.  The method allows the simultaneous determination of various parameters such as radius, masses, orbital period, luminosities and temperatures. The optimization method converged successfully for the initial parameters of orbital period P$_{i,o}= 3.9$ d, mass of the donor star M$_{i,d}=3.40$ M$_{\odot}$, mass of the gainer M$_{i,g}=3.18$ M$_{\odot}$ with a chi-square of $\Delta \chi^{2}_{0.95,8}= 0.212$ (Fig. 2).

We have compared the published values of the masses, radius, orbital period and radial velocities of V495 Cen by \citet{2018MNRAS.476.3039R} with our best model and the main feature presented is the great consistency between theoretical and observed values. We noted that the orbital period as a function of the mass of the primary component (donor) is relatively constant until to 2.5 M$_{\odot}$ and it changes abruptly around of 1.5 M$_{\odot}$ (Fig. 3). This suggests that the system is in a stage of change in its orbital period as the mass of the primary star varies. During the evolution of the system, we note that the orbital velocity of the primary component increases (Fig. 4, black line), reaches a peak around 150 km s$^{-1}$ and then decreases to about 106 km s$^{-1}$ for a mass of 0.91 m$_{\odot}$. The companion decreases its orbital velocity to 16 km s$^{-1}$ for a mass of 5.76 m$_{\odot}$ (Fig. 4, dashed red line). Also, we performed a study of the behavior of the radius as function of the mass for both stars (Fig. 5), that shows a small change in radius whereas the mass increases in the gainer star (dashed red line), possibly due to the rejuvenation process of this star produced by mass accretion. The primary component (black line) increases rapidly their size as its mass decreases. With respect to the mass transfer from the primary star toward the companion it proceeds in a quasi steady manner during most of time (Fig. 6). In summary, the theoretical model is consistent with the published results of V495 Cen.\\

\begin{figure}[]
	\begin{center}
		\includegraphics[width=7.2 cm,angle=0]{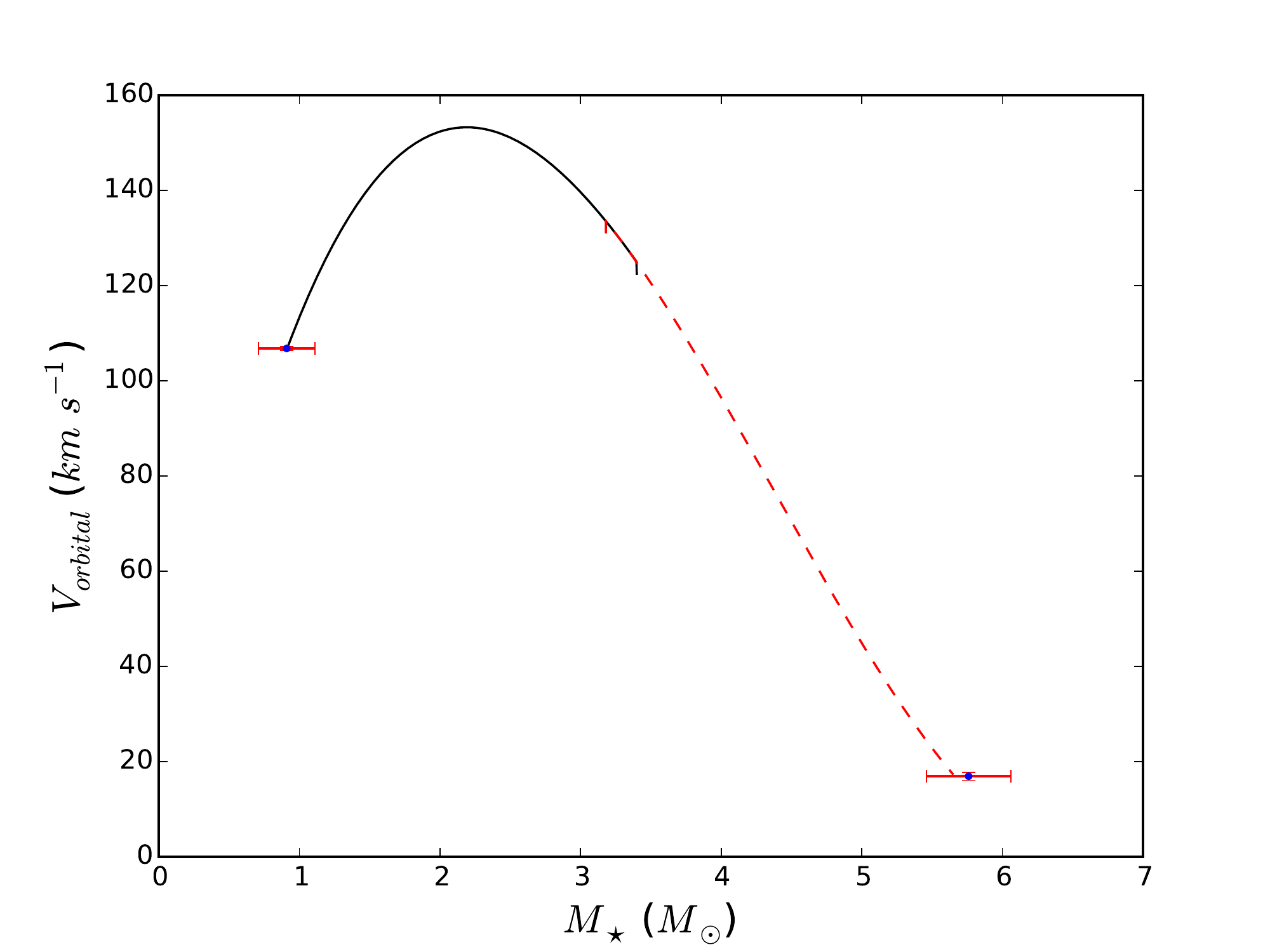}
	\end{center}
	\footnotesize
	{\bf Figure 4.} Orbital velocities curve for both components as a function of their masses. The primary component (black line) increases their orbital velocity during the loss mass process known as spin-up and later decrease reaching the current value. The secondary (dashed red line) component shows a diminutions of their radial velocity and a mass gain.
	\normalsize 
\end{figure}

\begin{figure}[h!]
	\begin{center}
		\includegraphics[width=7.2 cm,angle=0]{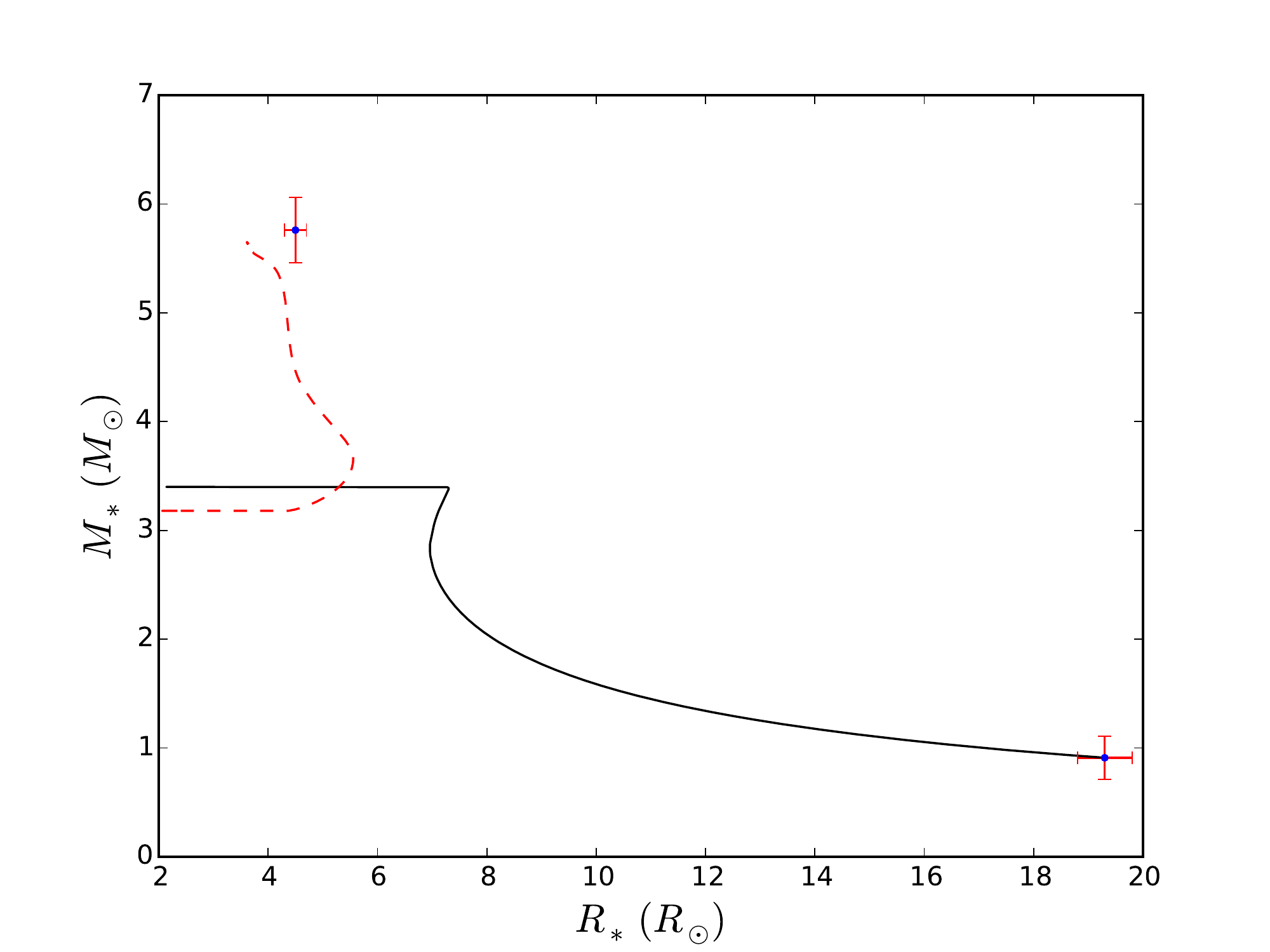}
	\end{center}
	\footnotesize
	{\bf Figure 5.} Schematic behavior of radius and mass for both stars. The gainer star (dashed red line) decreases its radius, while the donor star (black line) expands to fill the Roche lobe.
	\normalsize 
\end{figure}

\begin{figure}
	\begin{center}
		\includegraphics[width=7.2 cm,angle=0]{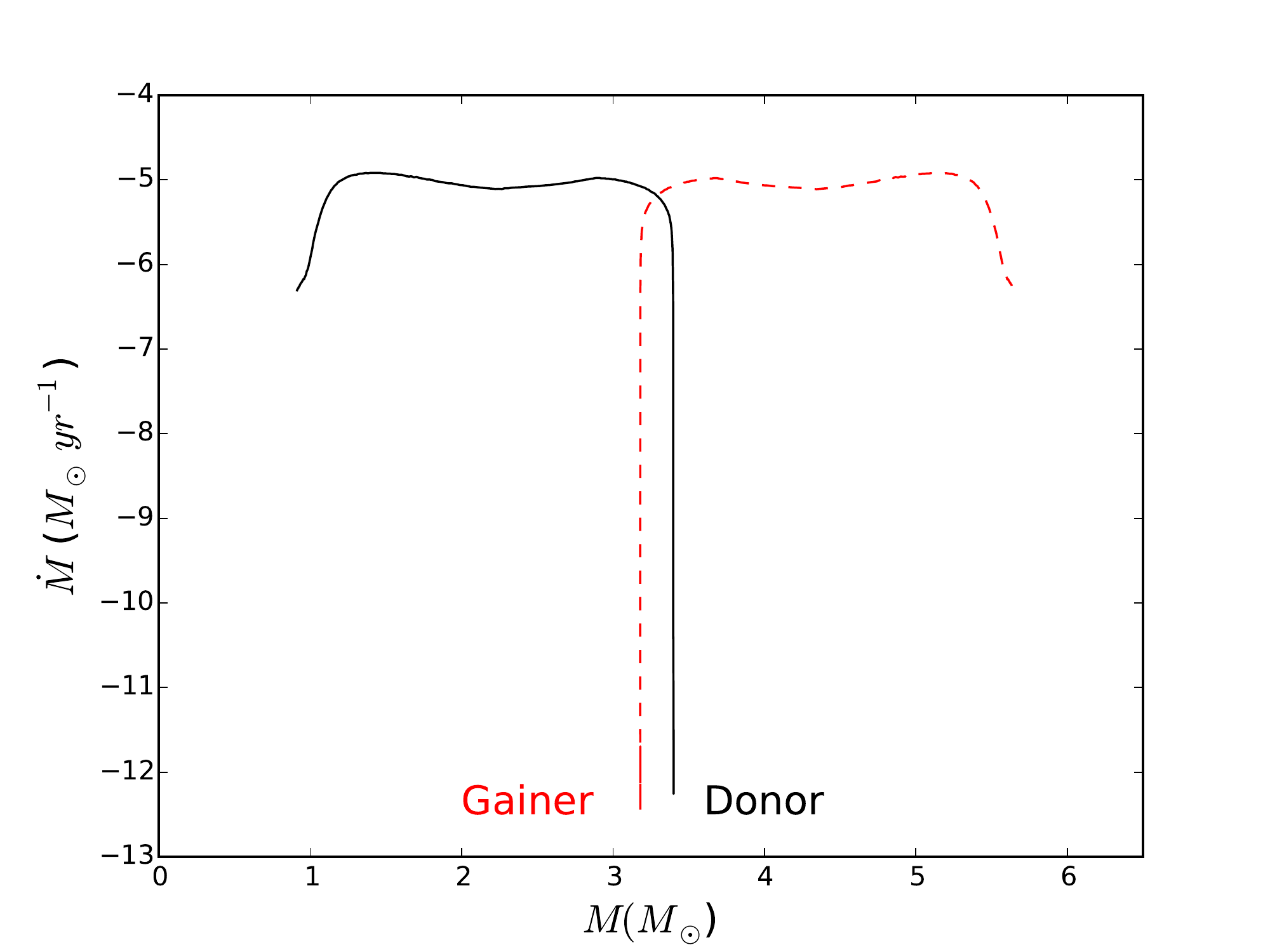}
	\end{center}
	\footnotesize
	{\bf Figure 6.} Theoretical variation of mass transfer until reaching the helium depletion for initial masses M$_{i,d}=3.40$ M$_{\odot}$, M$_{i,g}=3.18$ M$_{\odot}$ with initial orbital period P$_{i,o}=3.9$ d.
	\normalsize 
\end{figure}

\vskip 1cm
{\bf 3  EVOLUTION OF THE DPV 495 CEN IN THE HR DIAGRAM}\\

The evolution of both stars begins on the zero main sequence (ZAMS) with the hydrogen burning, labeled as A-point (Fig. 7). Both stars evolve similarly as they move, and the central helium abundances X$_{He,c}$ rises steadily as both stars fuse hydrogen into helium in their convective cores. During this phase, the primary star evolves, increasing its size and quickly depleting the central hydrogen, while the companion (gainer) preserves its initial volume, this can be seen in Fig 7. Once the donor star has exhausted the central hydrogen (B-point) at age 2.980$\times 10^{8}$ yr, it quickly start the hydrogen shell burning, causing an increase of its luminosity (C-point) and temperature, at age around 3.012 $\times 10^{8}$ yr. After this phase, the primary star decreases its luminosity by less than one order of magnitude and stabilizes its temperature, and starts (D-point) the optically thick mass transfer at age 3.0133 $\times 10^{8}$ yr, causing a rejuvenation of the companion (gainer star). The initial mass of the primary component decreases until reaching an equilibrium of both mass ($U_{1}$-point) where m$_{1}$=m$_{2}$, i.e., the mass ratio $q=1$ , at 3.0136 $\times 10^{8}$ yr. After mass ratio inversion, the donor star moves to the minimum value of the Roche lobe at 3.0142 $\times 10^{8}$ yr (E-point), arriving to the end of the phase of optically thick mass transfer ($U_{2}$-point) at 3.0159 $\times 10^{8}$ yr, where the donor luminosity drops about an order of magnitude, because the deeper layers of the donor need to expand as they adjust to the quickly decreasing mass. Finally, the donor star finishes the mass transfer stage (F-point) at the age 3.0268 $\times 10^{8}$ yr and then continues as a detached binary system. 

\begin{figure}
	\begin{center}
		\includegraphics[trim=1.2cm 0.2cm 0.2cm 0.8cm,clip,width=0.4\textwidth,angle=0]{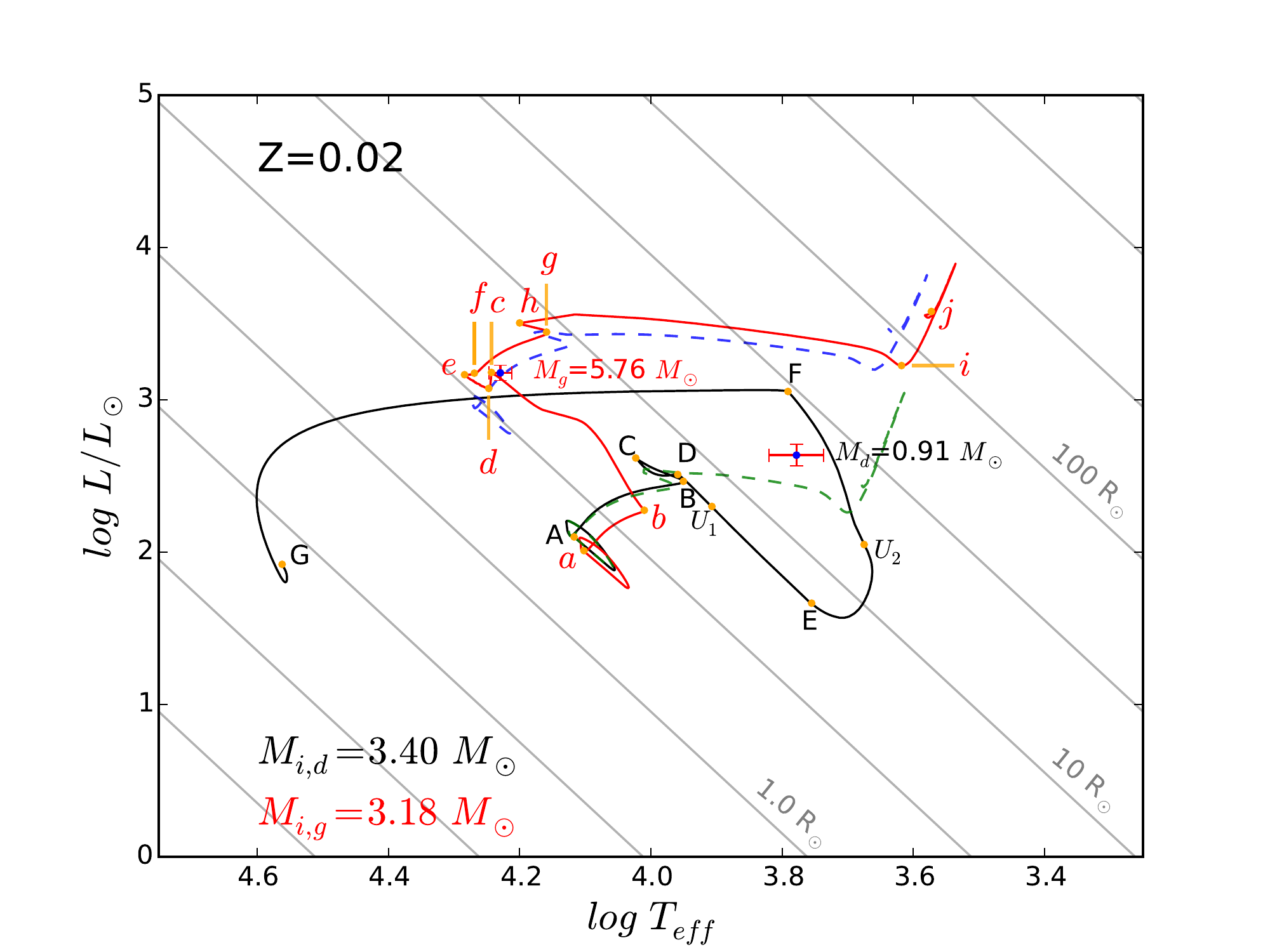}
	\end{center}
	\footnotesize
	{\bf Figure 7.} Evolutionary track of primary or donor star (black line) and its companion (red line). The important points of the evolution were labeled together to the initial masses. The dashed green line correspond to evolutionary track of a single star of initial mass M$_{i,s}=3.40$ M$_{\odot}$ using the same initial parameters of the donor star until the central helium depletion ($X_{H_{e,c}}<0.2$). The dashed blue line is the evolutive track for a single star of mass 5.76 M$_{\odot}$ with the similar characteristics of the gainer star.
	\normalsize 
\end{figure}

The stages C, D, $U_{1}$ of the donor star occur rapidly, causing a hook in the evolutionary track of the gainer star (Fig. 7, red line, $b$-point), preserving its volume and beginning a rejuvenation. The process that occurs at the D-point, illustrate the well-known fact that the change of the orbital period is completely bound to the mass transfer process (Fig. 9). This can seen in the Fig. 10, where we plot how the mass transfer changes as a function of time, also we indicate the processes of the onset of the mass transfer until the end of the optically thick mass transfer. The central density of the donor star is bound to the mass transfer, increasing gradually fast during the process, from the start of the mass transfer process until the mass inversion the central density increased $\sim$ 2.8 g cm$^{-3}$.

\begin{figure*}
	\begin{center}
		\includegraphics[width=8  cm,angle=0]{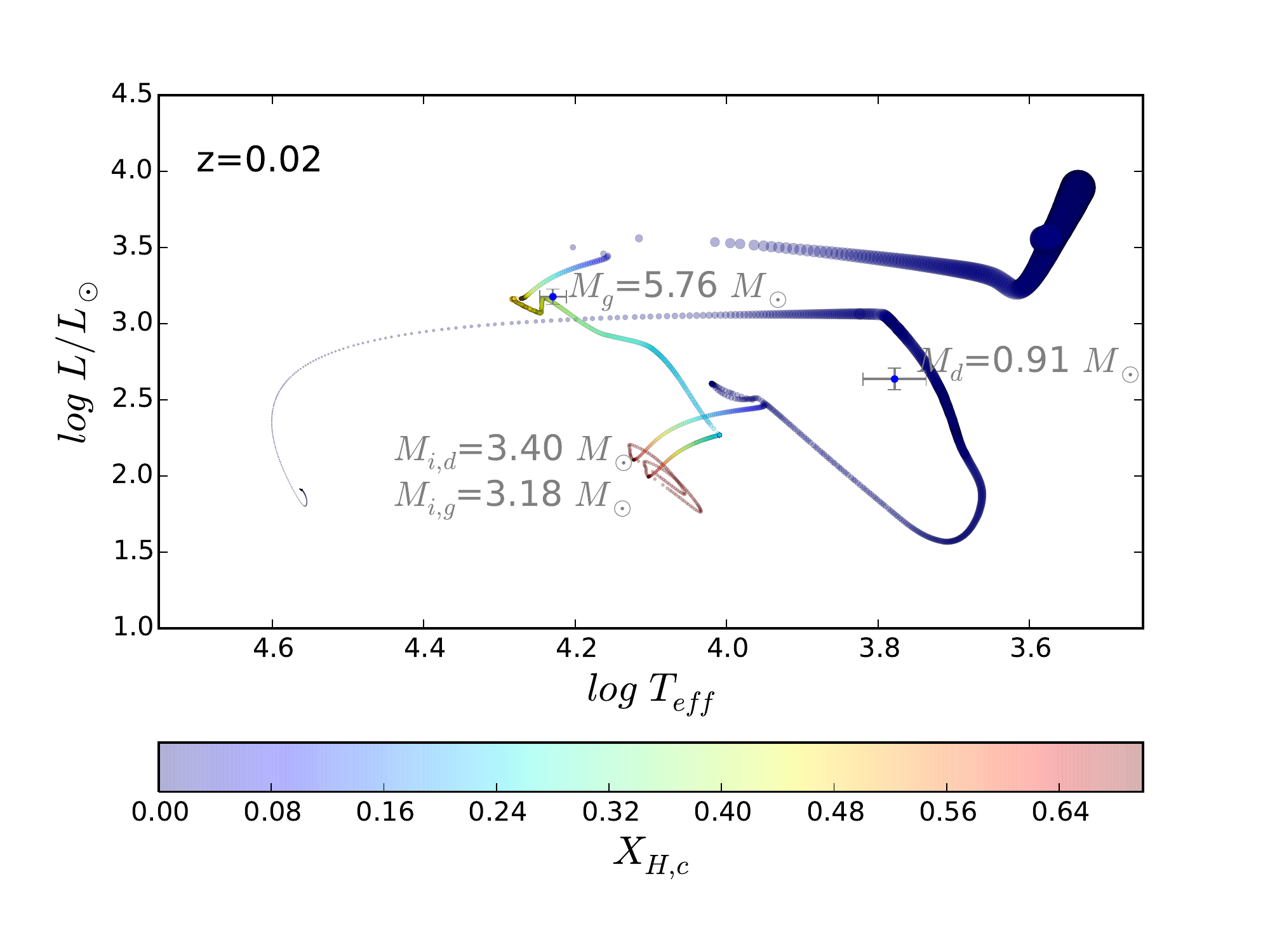}
		\includegraphics[width=8 cm,angle=0]{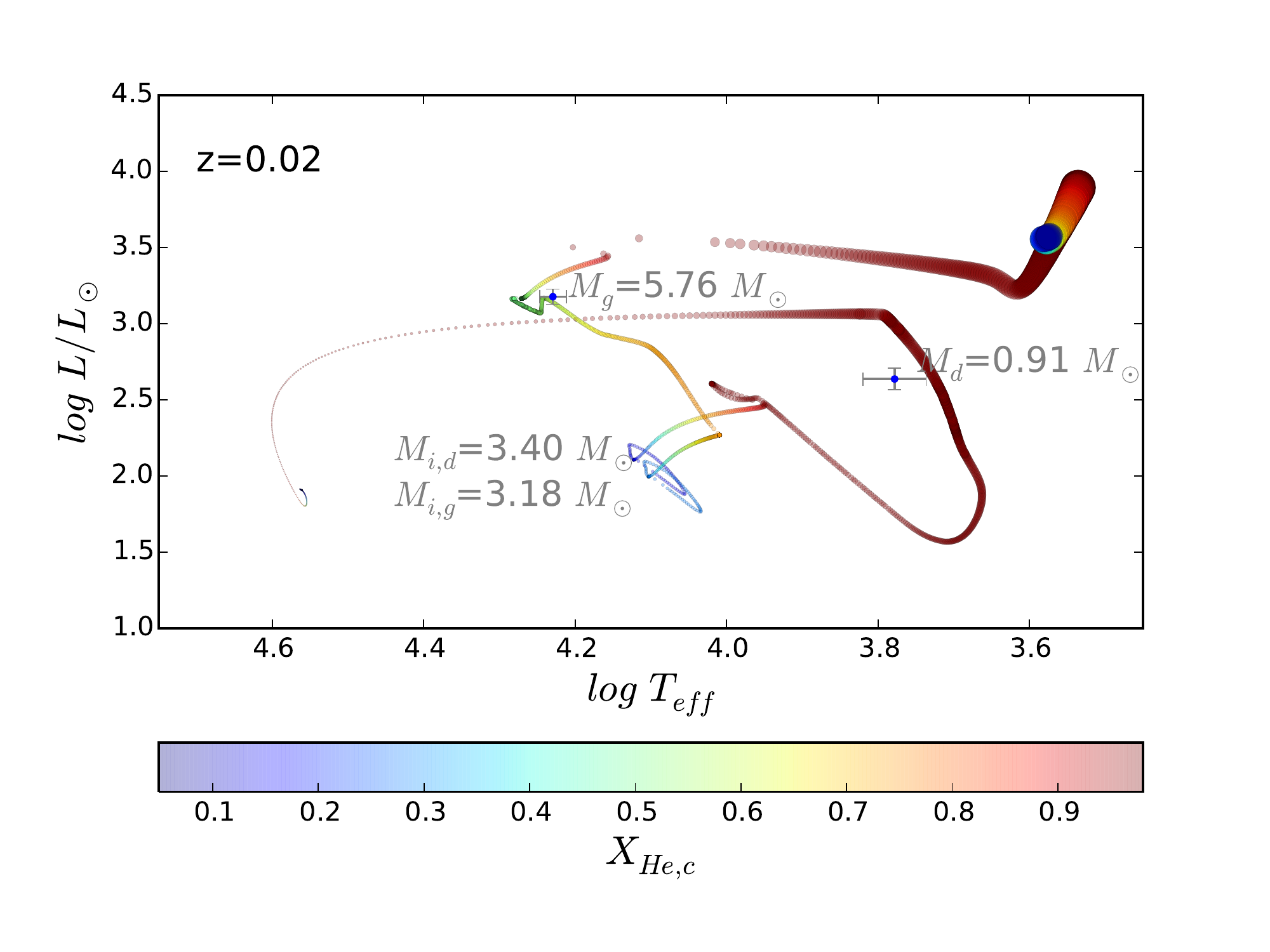}
	\end{center}
	\footnotesize
	{\bf Figure 8.} Hertzsprung-Russel (H-R) diagrams showing the binary evolution. The color bar shows the central hydrogen and helium mass fraction for both components and confirms a rejuvenation of the gainer retaining their size during great part of their lifetime, whereas that the donor star suffers a rapid evolution exhausting their helium during the mass transfer phase.
	\normalsize 
\end{figure*}

\begin{figure}[]
	\begin{center}
		\includegraphics[width=7.2 cm, angle=0 ]{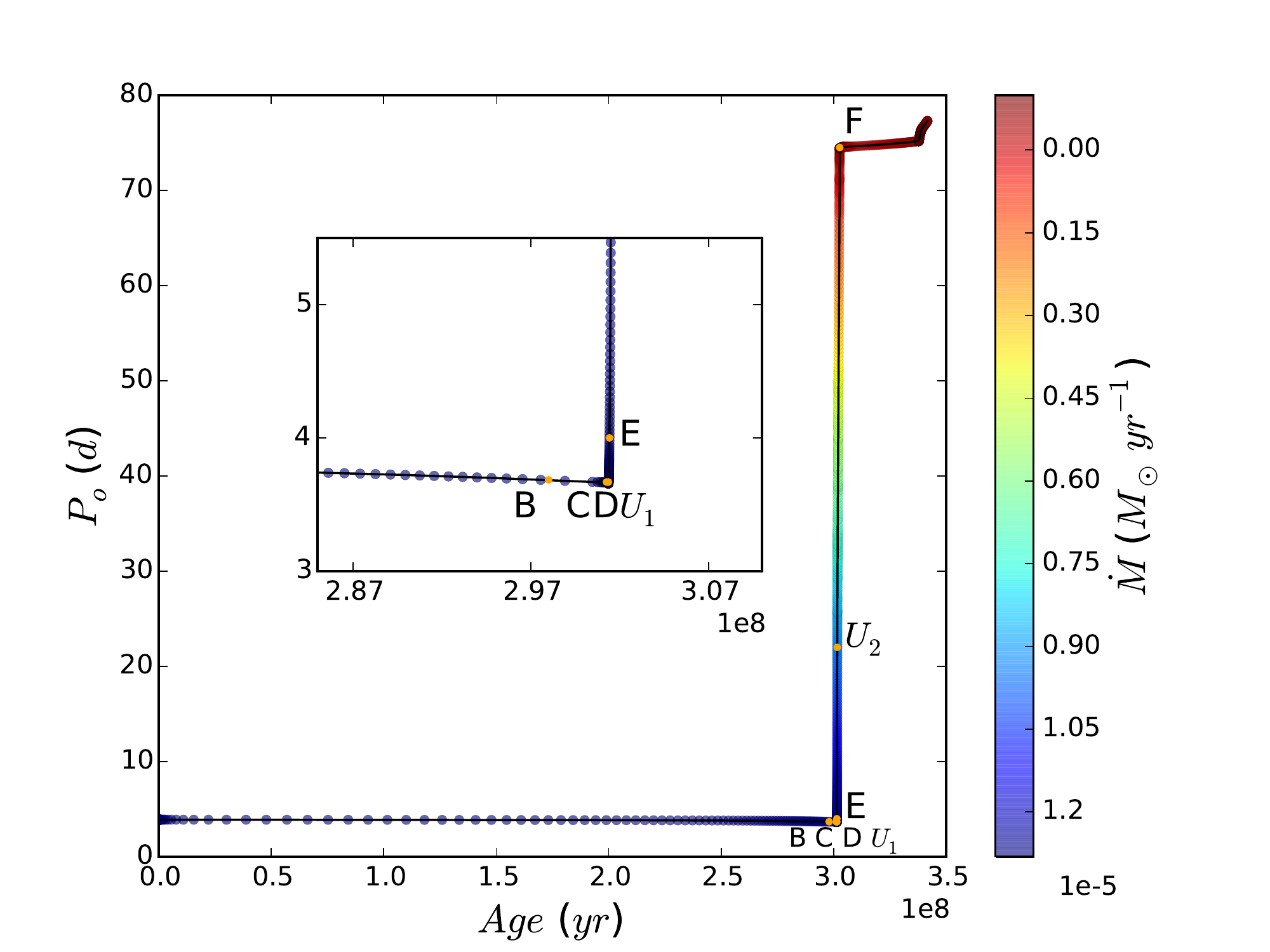}
	\end{center}
	\footnotesize
	{\bf Figure 9.} Change of the orbital period as a function of the time, labeled with the more important stage related to mass transfer and their change during the mass inversion.
	\normalsize 
\end{figure}

When the gainer begins the central hydrogen burning it reaches the \emph{b}-point. Then it starts to accrete mass, and rapidly recovers around of 17\% of its central hydrogen mass fraction from $X_{H,c}=0.240$ to $X_{H,c}=0.419$ until reaches the \emph{c}-point, causing the second hook in the evolutionary track. After the end of the optically thick mass transfer, the hot star stops its rejuvenation process but continue accreting matter, principally hydrogen from the disc, and begins a stage in which it tries to return to the main sequence, moving from the d-point to the e-point, while it continues the central hydrogen burning. Also, during the stages \emph{b} until \emph{e}, the temperature increases around 10,000 K and its luminosity an order of magnitude. Once that the maximum temperature and accreted mass has been reached (e-point), the new evolution of the gainer star begins in a sequence parallel to the main sequence for single stars, as if it were similar to a single star of M = 5.87 M$_{\odot}$. (Fig. 7, blue dashed line), in addition we added the single track evolution to a star of M = 3.40 M$_{\odot}$ in comparative with the donor star (Fig.7, green dashed line). \\

\begin{figure}
	\begin{center}
		\includegraphics[width=7.2 cm, angle=0 ]{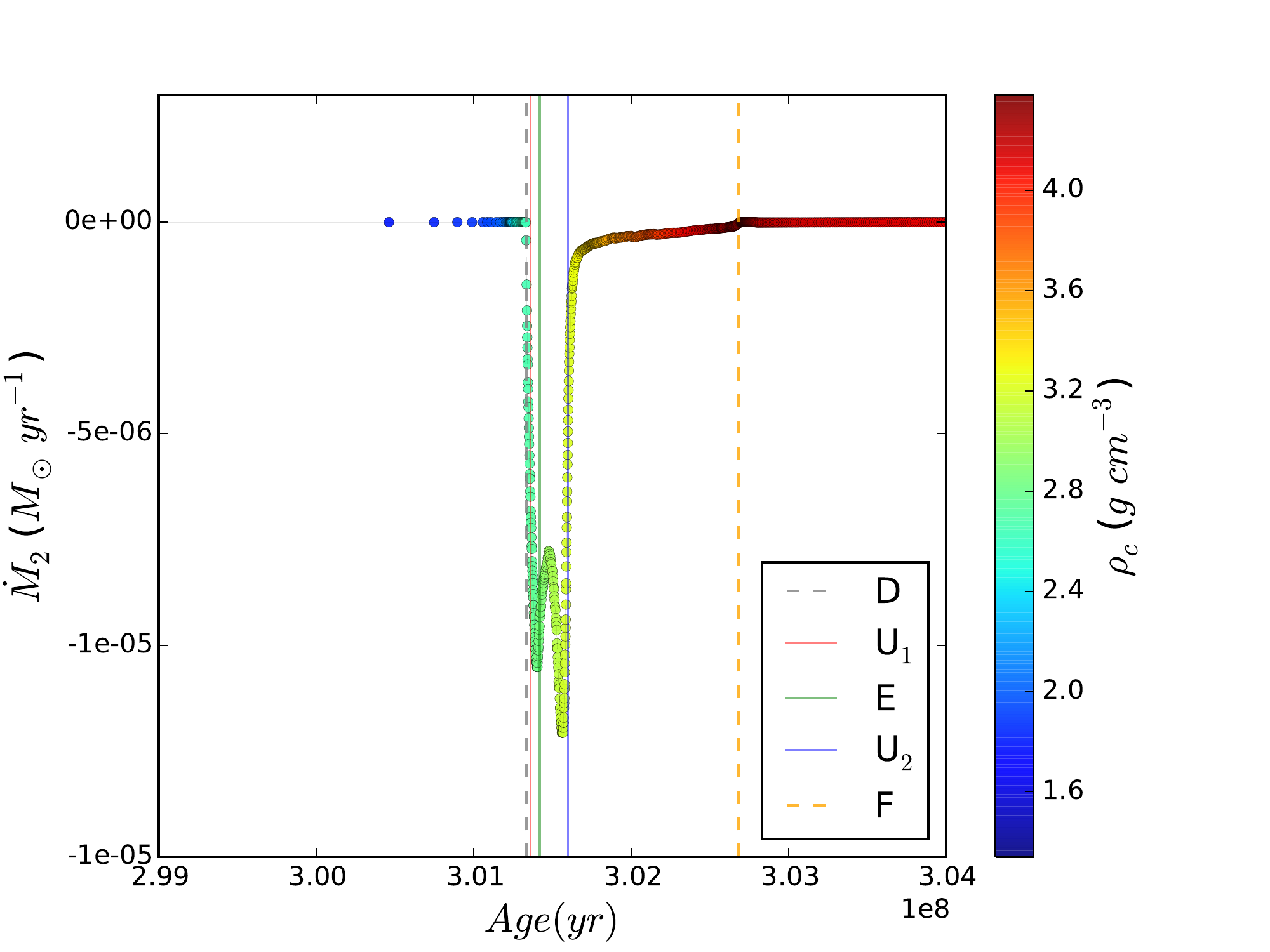}
	\end{center}
	\footnotesize
	{\bf Figure 10.} $\dot{M}_{d}$ curve for the best evolutionary model of the donor star. The vertical dashed line (black) indicates the optically thick mass transfer, the continuum the mass ratio q=1, i.e. m$_{1}$=m$_{2}$ and the continuum blue line corresponds to the end of the phase of optically thick mass transfer.
	\normalsize 
\end{figure}

\vskip 0.5cm
{\bf 4 STRUCTURAL CHANGES IN THE COMPONENTS OF V495 CEN DURING BINARY EVOLUTION}\\

For intermediate mass stars the theory of convective overshooting and semi convection is particularly complicated. Hence, it is important to emphasize  the three classical temperature gradients to constrain and understand the best model. The ratio between the natural logarithms of the derivate ($T$) and ($p$) is called $\nabla$ which is defined at any layer within of the star, while the radiative gradient $\nabla_{r}=(dln{T}/dln{P})_{r}$ represents the local heat flux by radiative transport. However, the gas can change adiabatically, therefore for a monatomic ideal gas the adiabatic gradient is $\nabla_{a}=(dln{T}/dln{P})_{s}$, where the $s$ indicate that the derivates are to be taken in the surrounding material. Therefore, if  $\nabla_{r}<\nabla_{a}$ in a certain layer, it is stable against convection, if $\nabla= \nabla_{r}$ all the heat is carried by radiation and finally if $\nabla_{r}>\nabla_{a}$ the layer is unstable to convection. We have constructed Kippenhahn diagrams to analyze the internal evolution of both stars with a stopping criterion when the donor reaches core helium depletion $X_{H_{e,c}}<0.2$(Fig. 11).

\begin{figure*}
	\begin{center}
		\includegraphics[trim=1cm 0.2cm 1cm 0.5cm,clip,width=0.31\textwidth,angle=0]{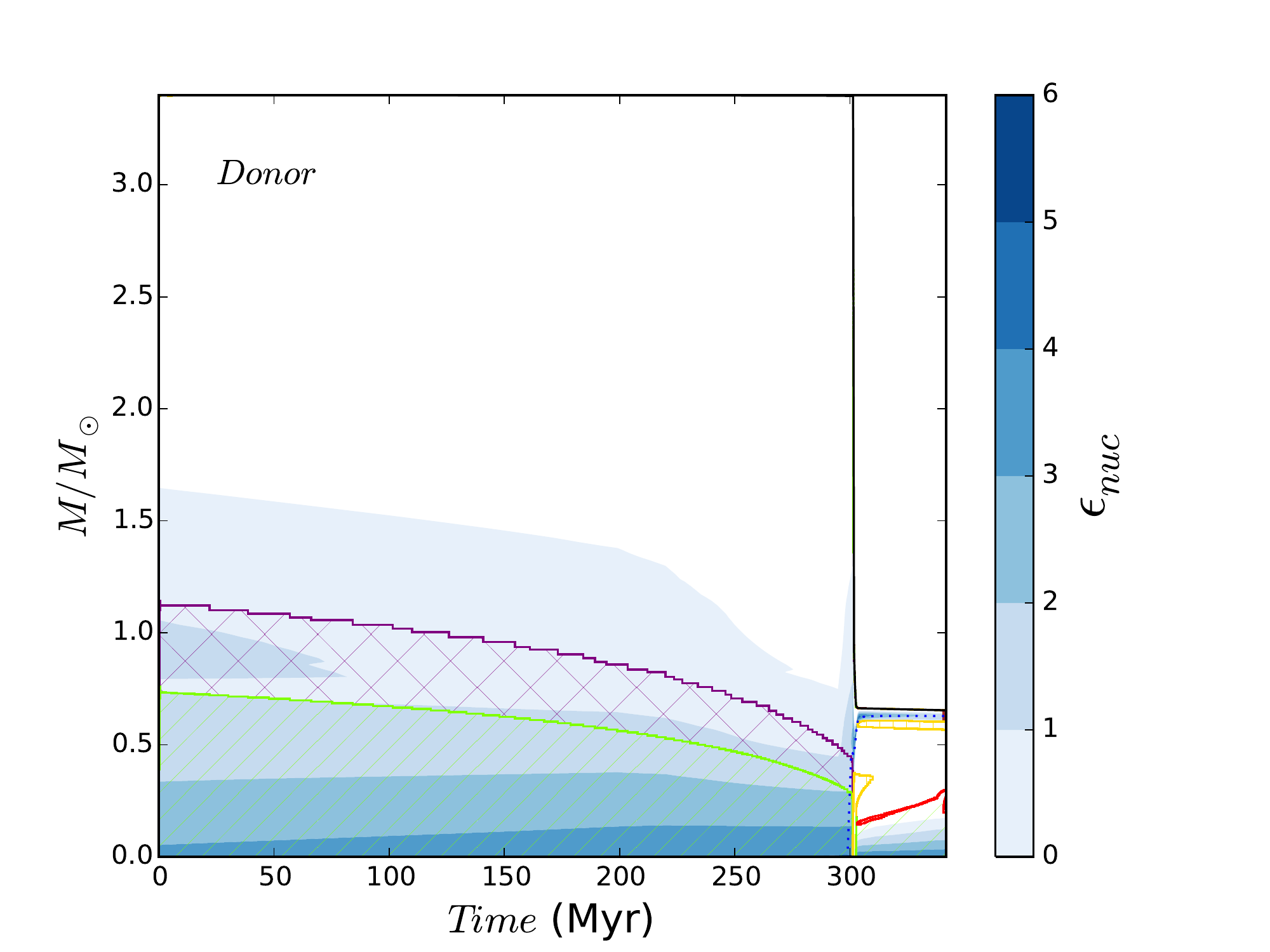}
		\includegraphics[trim=1cm 0.2cm 1cm 0.5cm,clip,width=0.31\textwidth,angle=0]{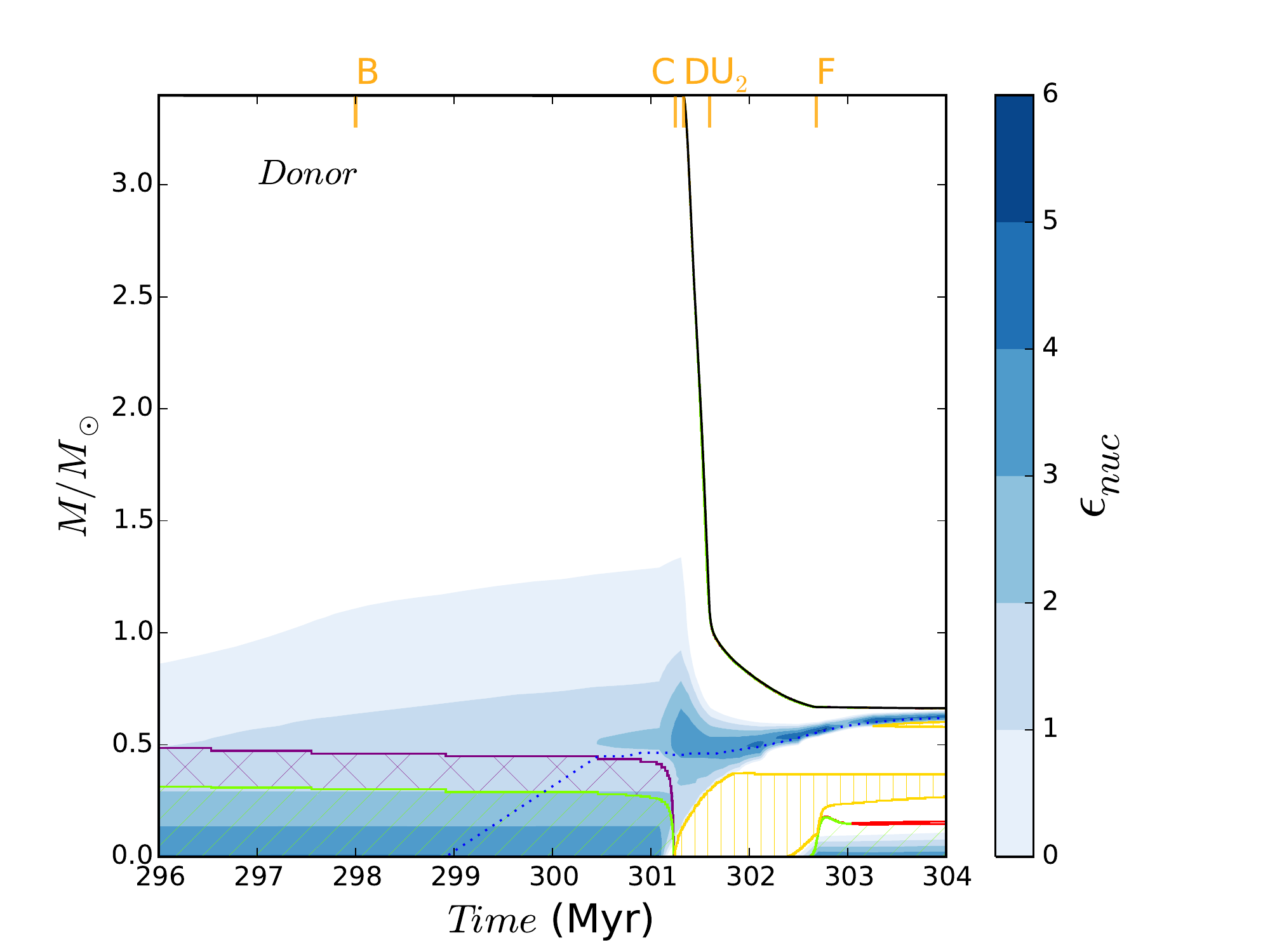}
		
		\includegraphics[trim=1cm 0.2cm 1cm 0.5cm,clip,width=0.31\textwidth,angle=0]{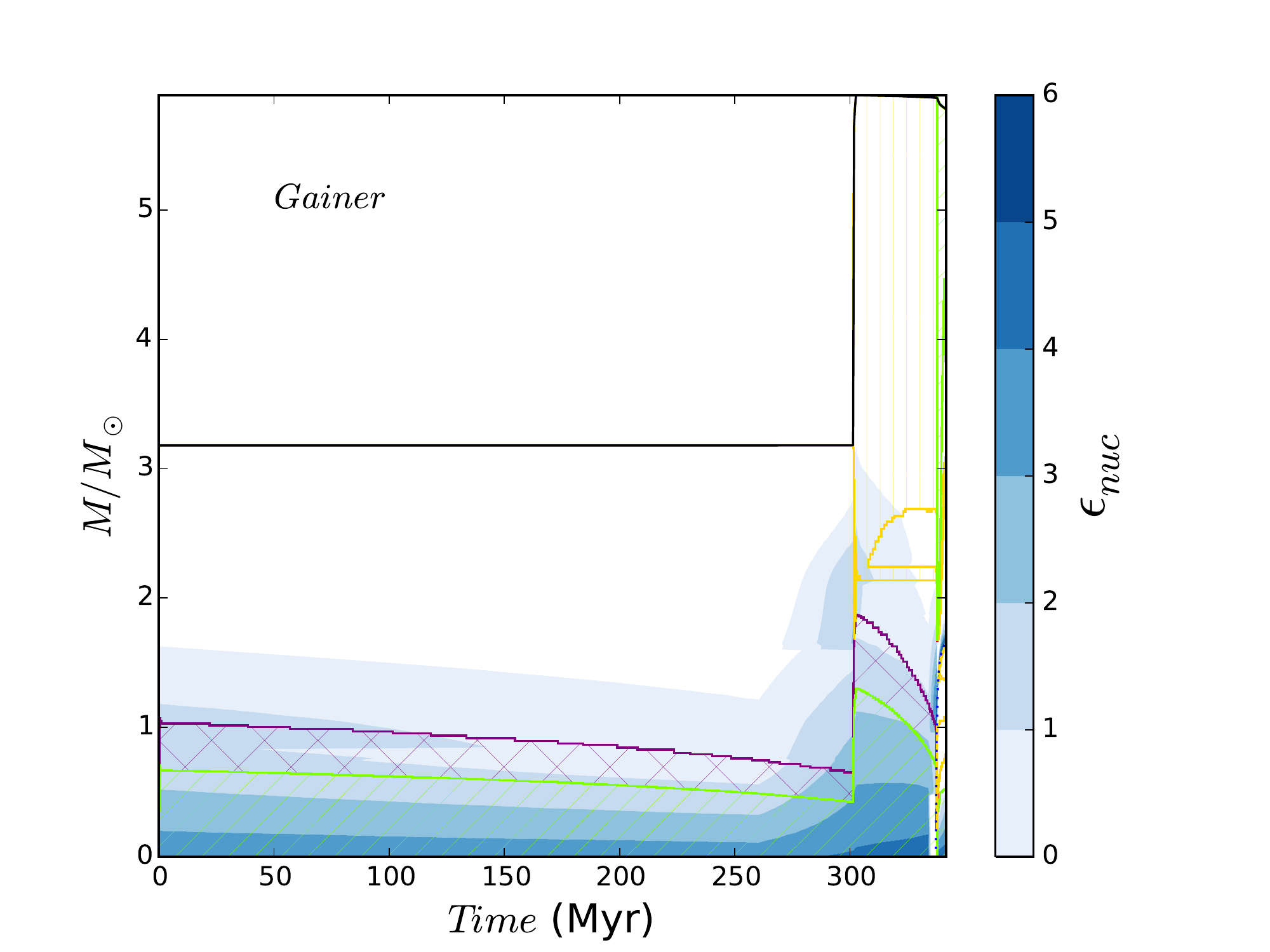}
		\includegraphics[trim=1cm 0.2cm 1cm 0.5cm,clip,width=0.31\textwidth,angle=0]{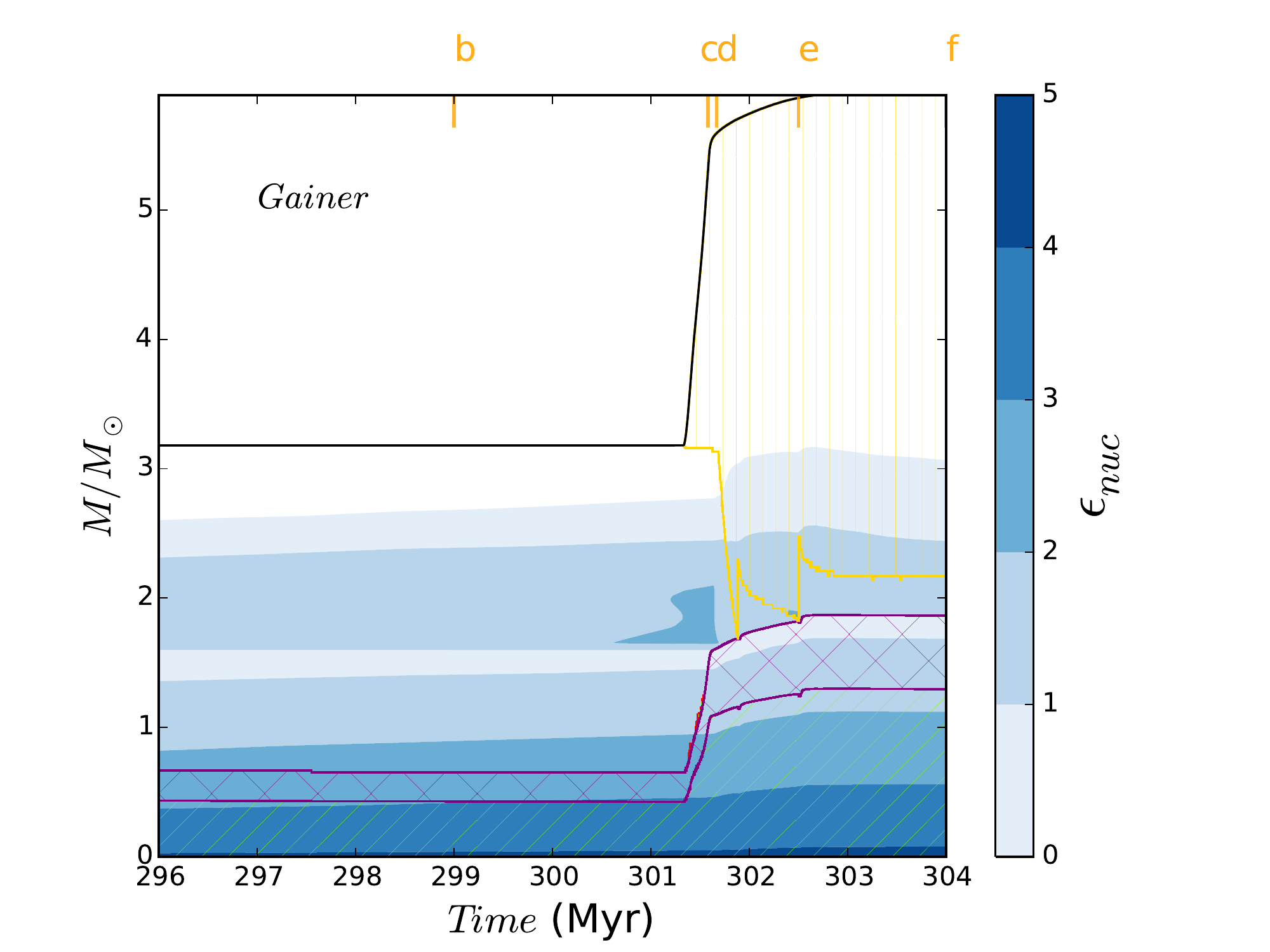}
	\end{center}
	\footnotesize
	{\bf Figure 11.} (Left) Kippenhahn diagrams showing the evolution of internal structure of both stars, accompanied (Right) by a zoom with some principal events labeled. (Up) diagram of donor star with initial mass M$_{d,i}$ = 3.40 M$_{\odot}$, (Down) diagram of gainer star with initial mass M$_{g,i}$ = 3.18 M$_{\odot}$. Both model stop when the donor star reaches core helium depletion $X_{H_{e,c}} < 0.2$. The x-axis give the age after ignition of hydrogen in units of Myr. The different layers are characterized by their values of M/M$_{\odot}$, convection mixing in hatched green, semi convection mixing in red, overshooting mixing in crosshatched purple, thermohaline mixing in hatched yellow, the solid black line shows the surface of each star and brown zone corresponds to rotational mixing.
	\normalsize 
\end{figure*}

The donor star begins from the zero age main sequence (zams) with hydrogen burning and a convective core (hatched green) around of $\sim$ 0.75 M$_{\odot}$ which slowly decreases around of $45\%$ of its central mass at C-point, previously to the moment of beginning the mass transfer process. Practically, half of the energy generation rate is produced inside of the convective core around of 0.32 M$_{\odot}$. After the convective core, we noted an overshooting convection region (crosshatched purple) where the material is carried from an unstable zone to stratified region (Fig. 11, up/left), this process causes helium core mass increase after Hydrogen depletion at 299 Myr steadily (Fig. 11, up/right, blue dots), later it enters a resting stage during 1.05 Myr whose stage corresponds to the increase of luminosity and temperature (C-point) until the end of optically thick mass transfer (U$_{2}$-point), later the core mass at the end of the main sequence becomes larger than would be expected in single stars, followed by an uncontrolled decrease of the total mass of the star due to the mass transfer. After to the mass transfer, appears a mechanism which governs the photospheric composition: the double diffuse instability is observed as a kind of elongated fingers, caused by the molecular weight inversion created by the $^{3}$He and $^{4}$He \citep{2007A&A...476L..29C} in the external layer of hydrogen burning, called thermohaline mixing (Fig. 11, yellow line). Also we detected that the convective core is governed by a slim layer of semi convective mixing (red line) and for a short time the donor star presents an increase of its internal rotational velocity during the helium depletion producing new stream in the stellar matter conducing the matter from the center to the surface and  return to the center in a short time (Fig. 12, brown square) at 340.4 Myr, mixing the matter of the core with the envelope of the star.

The evolution of the hot companion or gainer star is more interesting since the external layer of the star is governed by a thermohaline mixing process, which shows a convective core of 0.67 M$_{\odot}$ where around of 10\% of the core presents an overshooting mixing process and later 301.3 Myr has an increase in the energy production rate. After mass transfer, the convective core increases its mass reaching a maximum of 1.30 M$_{\odot}$, moving the overshooting mixing process the same proportion and increasing its energy generation deeper in the core. The thermohaline region shows an anonymous mixing later at 300 Myr which is controlled by convective layers of the gainer. The main evolutionary stages for both stars are given in Table 1.\\

\begin{table*}
	\footnotesize
	\caption{Evolutive stages for DPV V495 Cen until $^{4}$He depletion of the donor star and their main feautures.}
	\normalsize
	\[
	\resizebox{16.0cm}{5.2cm}{$
		\begin{array}{llrrrrl}
		\hline
		\hline
		\noalign{\smallskip}
		\textrm{}  		& \textrm{Stage} 	& \textrm{Age (Myr)}		& \textrm{M (M$_{\odot}$)}& \textrm{R$_{\odot}$} & \textrm{log T (K)} & \textrm{Ev. Process}\\
		\hline
		\textrm{Donor}  & \textrm{}		 	& \textrm{}			& \textrm{}			& \textrm{}	 		& \textrm{} 	& \textrm{}\\
		\textrm{}  		& \textrm{Z.A.M.S.}	& \textrm{0.000} 	& \textrm{3.400}	& \textrm{2.183}	& \textrm{4.116}& \textrm{\tiny{Zero Age Main Sequence, Hydrogen burning.}  }\\
		\textrm{}  		&\textrm{T.A.M.S.} 	&\textrm{52.370}  	&\textrm{3.399}  	&\textrm{2.359}  	&\textrm{4.111}	& \textrm{\tiny{Terminal age main sequence}}\\
		\textrm{}  		& \textrm{B}  		& \textrm{298.000} 	& \textrm{3.397}	& \textrm{7.201}	& \textrm{3.949}& \textrm{\tiny{$^{1}$H depletion (Sub. giant).}}\\
		\textrm{}  		& \textrm{C}  		& \textrm{301.250} 	& \textrm{3.397}	& \textrm{6.088}	& \textrm{4.018}& \textrm{\tiny{Increase of luminosity and temperature.}  }\\
		\textrm{}  		& \textrm{D}  		& \textrm{301.335} 	& \textrm{3.395}	& \textrm{7.287}	& \textrm{3.955}& \textrm{\tiny{Mass transfer.}}\\
		\textrm{}  		& \textrm{U$_{1}$}  & \textrm{301.360} 	& \textrm{3.288}	& \textrm{7.214}	& \textrm{3.909}& \textrm{\tiny{Inversion masses.}}\\
		\textrm{}  		& \textrm{E}  		& \textrm{301.420} 	& \textrm{2.732}	& \textrm{6.980}	& \textrm{3.755}& \textrm{\tiny{Minimun value Roche lobe (Red giant).}}\\
		\textrm{}  		& \textrm{U$_{2}$}  & \textrm{301.598} 	& \textrm{1.089}	& \textrm{15.376}	& \textrm{3.670}& \textrm{\tiny{End of optically thick mass transfer.} }\\
		\textrm{}  		& \textrm{F}  		& \textrm{302.681} 	& \textrm{0.668}	& \textrm{29.269}	& \textrm{3.792}& \textrm{\tiny{End of mass transfer stage.} }\\
		\textrm{}  		& \textrm{G}  		& \textrm{341.701} 	& \textrm{0.653}	& \textrm{0.225}	& \textrm{4.563}& \textrm{\tiny{$^{4}$He depletion (W. D.).} }\\
		\hline
		\textrm{Gainer} & \textrm{} 	   & \textrm{}				& \textrm{}			& \textrm{}	 		& \textrm{}		& \textrm{}\\
		\textrm{}  		& \textrm{Z.A.M.S.}& \textrm{0.000}		& \textrm{3.180}	& \textrm{2.100}	& \textrm{4.101}& \textrm{\tiny{Zero Age Main Sequence}}\\
		\textrm{}  		&\textrm{T.A.M.S.} &\textrm{52.370}  	&\textrm{3.179}  	&\textrm{2.236}  	&\textrm{4.093} & \textrm{\tiny{Terminal age main sequence}}\\
		\textrm{}  		& \textrm{b}  	& \textrm{299.000}		& \textrm{3.179}	& \textrm{4.299}	& \textrm{4.011}& \textrm{\tiny{$^{1}$H Enrichment}}\\
		\textrm{}  		& \textrm{c}  	& \textrm{301.580} 		& \textrm{5.337}	& \textrm{4.203}	& \textrm{4.241}& \textrm{\tiny{End of rejuvenation stage} }\\
		\textrm{}  		& \textrm{d}  	& \textrm{301.670} 		& \textrm{5.595}	& \textrm{3.671}	& \textrm{4.246}& \textrm{\tiny{Last hydrogen accretion from inner of the disc} }\\
		\textrm{}  		& \textrm{e}  	& \textrm{302.500} 		& \textrm{5.866}	& \textrm{3.496}	& \textrm{4.279}& \textrm{\tiny{Main Sequence relocation} }\\
		\textrm{}  		& \textrm{f}  	& \textrm{304.000} 		& \textrm{5.886}	& \textrm{3.773}	& \textrm{4.266}& \textrm{\tiny{$^{1}$H re-burning}}\\
		\textrm{}  		& \textrm{g}  	& \textrm{335.000} 		& \textrm{5.870}	& \textrm{7.924}	& \textrm{4.166}& \textrm{\tiny{Sub giant} }\\
		\textrm{}  		& \textrm{h}  	& \textrm{337.600} 		& \textrm{5.867}	& \textrm{7.380}	& \textrm{4.202}& \textrm{\tiny{Early horizontal branch} }\\
		\textrm{} 		& \textrm{i} 	& \textrm{337.826}		& \textrm{5.867}	& \textrm{78.509}	& \textrm{3.620}& \textrm{\tiny{Red clump}}\\
		\textrm{}  		& \textrm{j}  	& \textrm{341.701} 	& \textrm{5.778}	& \textrm{145.470}	& \textrm{3.573}& \textrm{\tiny{$^{4}$He depletion (A. G. branch)} }\\
		\hline
		\hline
		\end{array}
		$}
	\]
\end{table*}

\vskip 0.5cm
{\bf 5 STELLAR DYNAMO IN THE DONOR STAR}\\

In this section we have developed a brief analysis about the effects of the magnetic fields in rotating stars. Typically the magnetic field generation by differential rotation has been regarded as a process operating in the convective zone.  However, the magnetic fields can be created in stratified layers in differentially rotating stars. Also the convection is not really necessary for a dynamo process to operate if the toroidal field replaces the role of convection as a result of magnetic instability. Hence, the differential rotation is an azimuthal amplification mechanism which stretches the field lined and forms a toroidal field \citep{2002A&A...381..923S}. We have analyzed the magnetic fields generated by the Tayler-Spruit dynamo during three different stages of the system : \\

\begin{itemize}
\item The first stage corresponds to hydrogen depletion to 298.000 Myr (M$_{d}$=3.397 M$_{\odot}$, M$_{g}$=3.179 M$_{\odot}$).
\item During the mass transfer, i.e. up to 301.335 Myr (M$_{d}$=3.395 M$_{\odot}$, M$_{g}$=3.181 M$_{\odot}$).
\item The current stage of the system to 301.763 Myr (M$_{d}$=0.913 M$_{\odot}$, M$_{g}$=5.651 M$_{\odot}$).\\
\end{itemize}

\noindent
We notice that the Eulerian diffusion coefficient for mixing $D_{Eulerian}$ (Fig. 12, yellow line) decreases when it gets close to the limit of the convective mixing zone and remains constant within the overshooting mixing zone at 298.000 Myr. Out of both zones this mechanism stops abruptly, and the elements that exhibit greater diffusivity outside this zone will be transported by advection, i.e., will be transported by the velocity of the fluid causing that the diffusion coefficient gradually increase again until almost reaches the original values at the most outer layer of the star. It is possible that this events called advection has been able to mechanically grab a hold of electric charges and force currents to flow, which would leads to generation of seed magnetic field and the presence of magnetic fields outside the convective and overshooting mixing zones (Fig. 12, Top), as proposed by \citet{1950ZNatA...5...65B}. Thus, the advection has an important role when we deal with about stellar dynamo in the donor star.

Once both magnetic fields are generated, these move towards the convective core and we notice that the toroidal magnetic field dominates the poloidal magnetic field during all the evolution track.  Since now the energy is transported by radiation (radiative zone), the ST diffusion coefficient becomes important,  allowing the mixing to occur in the radial direction. This coefficient dominates over the Eulerian diffusion coefficient and is related to the poloidal magnetic field in their first moments. The greater variation of the toroidal and poloidal components occurs previous to the mass transfer at the upper limit of the overshooting mixing zone from 0.43 to 1.23 M$_{\odot}$ (Fig. 12, center), causing a decrease of the diffusion coefficient (cm$^{2}$ s$^{-1}$) within of the convective core around of 11.8 units but allowing an increase of the overshooting mixing zone in almost  7 units too. We notice that since the hydrogen depletion until the current stage of V495 Cen, the magnetic fields toroidal and poloidal  increased in 1.5 and 1.3 dex respectively (Fig. 12, Bottom). The expected toroidal magnetic field strength at the surface of the donor should be between 0 to 100 G and, while the poloidal between 0 to 2000 G.  Theses intervals were estimated agree to the magnetic field strengths found by \citet{2013A&A...554A.134L} and \citet{2005A&A...434.1029V} in evolve and late type stars. We noted that the obtained values close to the surface for the toroidal magnetic field $B_{\phi} \sim 15$ G and to poloidal magnetic field B$_{r} \sim 1900$ G are coherent with the expected values.. Also the diffusivity of some elements inhibits the presence of magnetic fields and the appearance of these could be related to the advection, generating magnetic seeds and later the respective magnetic fields.\\

\begin{figure}[h!]
	\begin{center}
		\includegraphics[width=7.5  cm,angle=0]{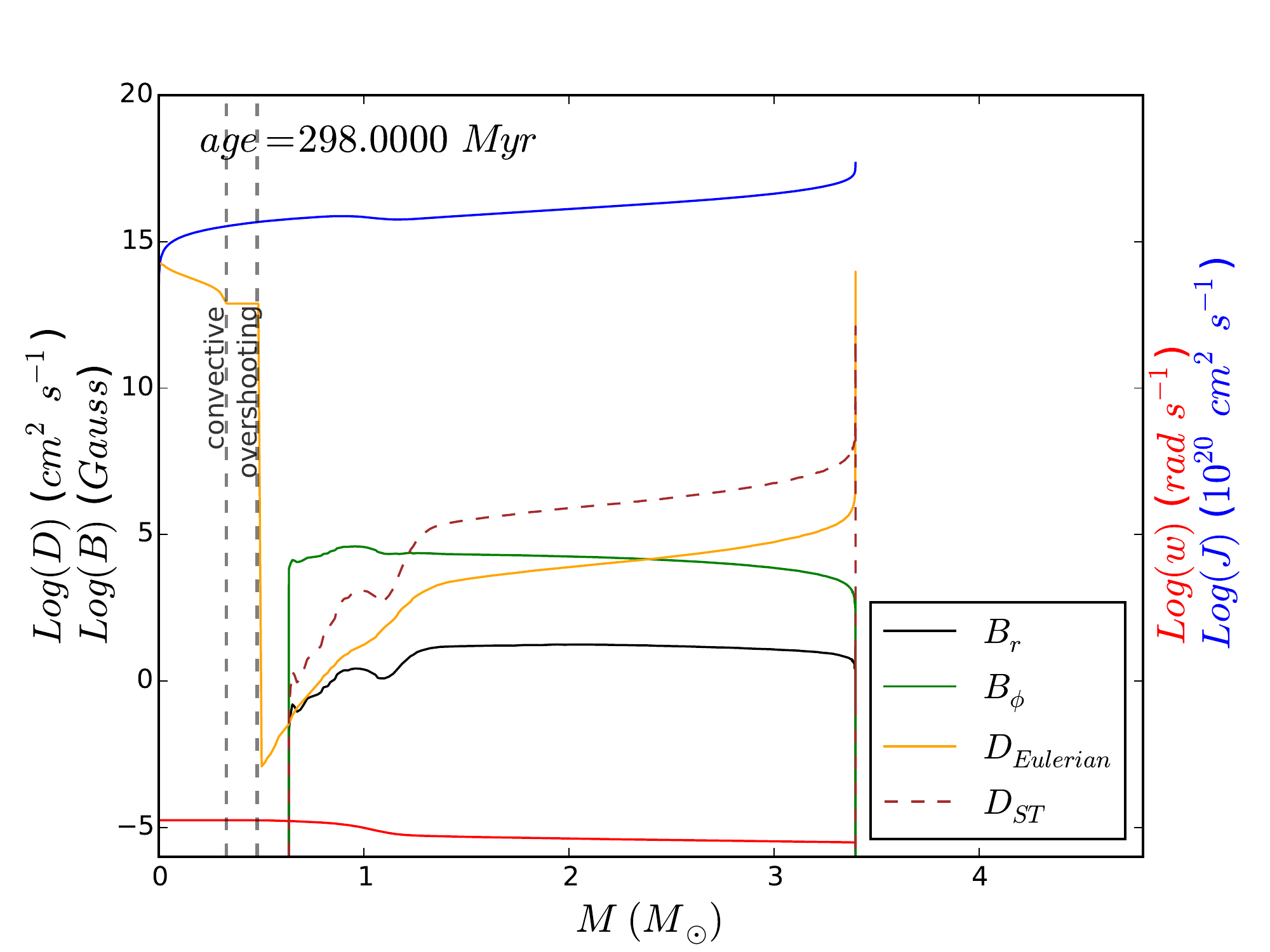}
		\includegraphics[width=7.5  cm,angle=0]{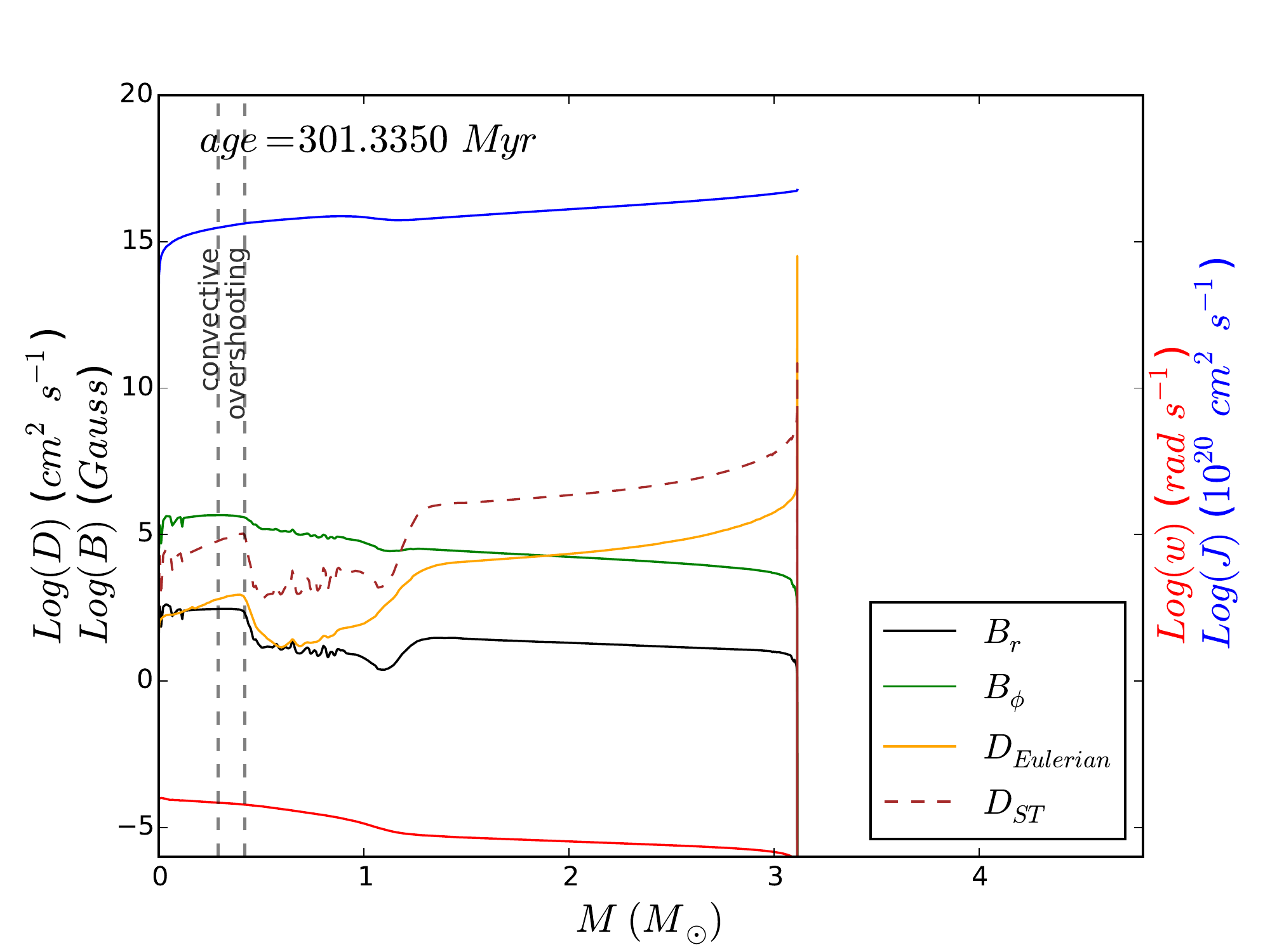}
		\includegraphics[width=7.5  cm,angle=0]{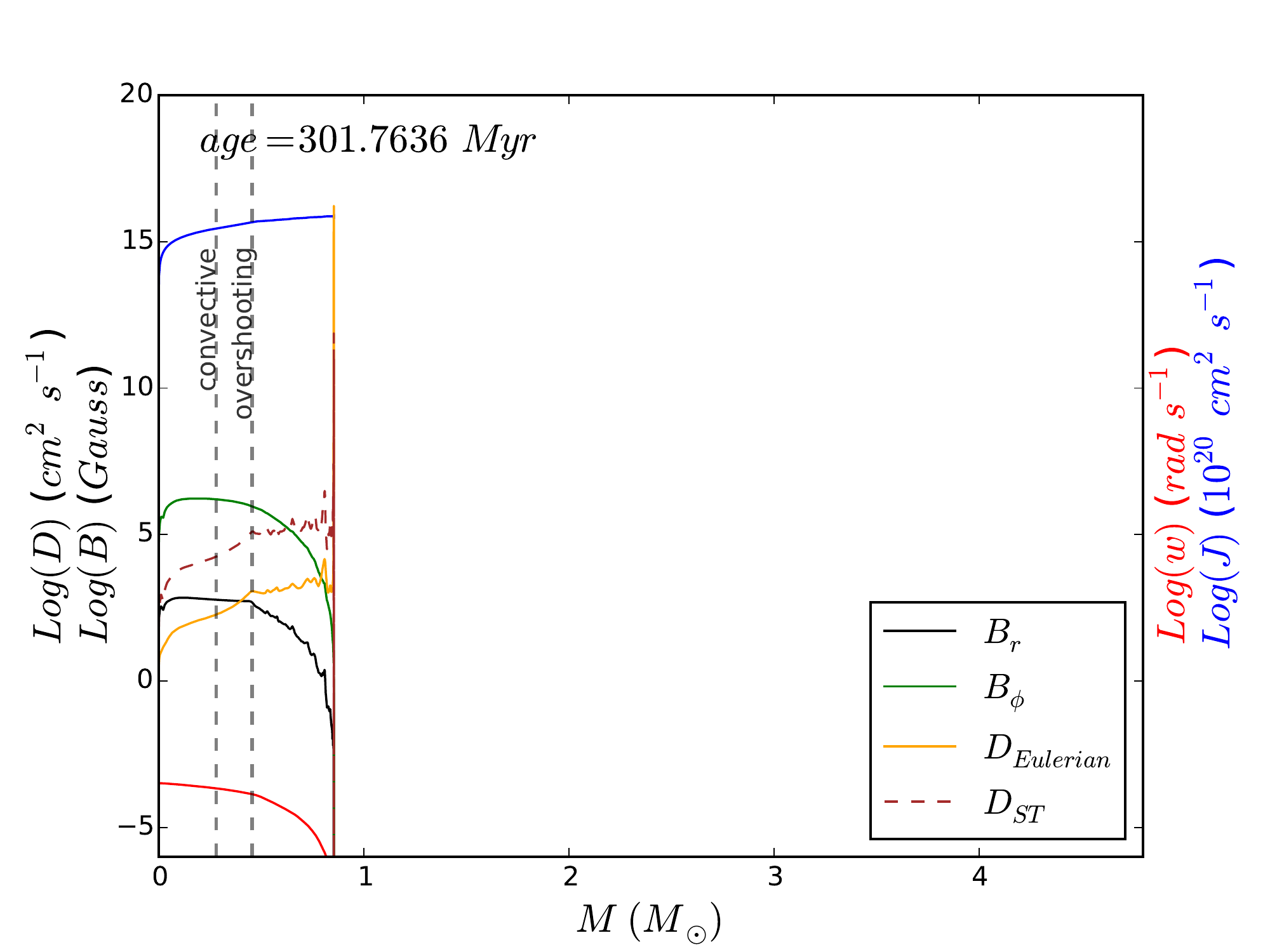}
	\end{center}
	\footnotesize
	{\bf Figure 12.} Three profiles of the magnetic fields generated by the Tayler-Spruit dynamo in the poloidal (radial) and toroidal (azimuthal) components to the donor star. From top to bottom the profiles are hydrogen depletion, mass transfer and current stage of the system. The Eulerian diffusion coefficient (yellow) out of the convective zone is stopping and allow to the Spruit-Tayler (ST) diffusion coefficient (Brown-dashed) work within radiative zone. The poloidal magnetic field (black) is modeled by the diffusion ST coefficient, while the toroidal magnetic field (green) is dominant as is expected. The angular momentum and velocity are represented in blue and red, respectively.
	\normalsize 
\end{figure}

\vskip 0.5cm
{\bf 6 DISCUSSION}\\

The Spruit-Tayler dynamo model implemented in {\textsc{mesa}}, shows how magnetic fields can be produced in a layer outside the convection and overshooting zones. At the same time the rotation and internal movement of the star contribute as an energy source for the process of generating magnetic fields.
We know that using a particular dynamo model does not guarantee us to predict the second photometric variability of the DPVs. However, we understood that the presence of magnetic fields within the donor is definitely justified. Therefore, the try to forecast the magnetic evolution of the DPV using the characteristics of a almost conservative mass transport. The differential rotation can be produced by a torque like a stellar wind or by internal evolution of the star is in certain way correct, and it is the start to produce the first predictive models for DPVs \citep{2017A&A...602A.109S}. 

Unfortunately, still we do not know what type of dynamos is most appropriate for these systems, without forgetting the mechanisms that drive their duration and variation of amplitude in the long cycle. Thus it leaves us with the following open question: how do we accurately predict a dynamo solution which describes the second photometric variability of the DPVs?. As a potentially useful insight, we notice that the intensity of the magnetic fields at the three different stages computed with our model for V495 Centauri are in agree to the magnetic field strengths found by \citet{2013A&A...554A.134L} and \citet{2005A&A...434.1029V} in evolve and late type stars.

The main uncertainties associated to our simulations are associated to the initial stellar parameters like masses, orbital period and luminosities. We intended different combinations of these parameters to fit the observed configuration of V 495 Centauri. As the	{\textsc{mesa}} code follows standard and well established stellar evolution theory, the predominant errors comes from the observed masses, radii and luminosities, being the orbital period the best determined parameter. On the other hand, the different dynamo models included in the {\textsc{mesa}} code follow physical prescriptions with relative importance depending on the stellar evolutionary stage. The basic assumption here is the veracity of these prescriptions. The errors associated to the dynamo in {\textsc{mesa}} are less important compared with the errors associated to the overall evolutionary track (controlled basically by nuclear evolution) and far less than those related to the observed data.

\vskip 0.5cm
{\bf 7 CONCLUSIONS}\\

In this work we have presented a detailed view of the evolutionary stages of the close interacting binary V495 Cen, following the evolution with  the code {\textsc{mesa}}. We also compared the results of this evolutionary route with the published parameters of the binary, finding the best representative model. The main results of our research are: 

\begin{itemize}

\item The model suggests that the orbital period is stable during the evolution as single stars, until the primary star fills the Roche lobe, causing an abrupt variation and increase during the mass transfer process.

\item The Donor star shows a maximum of its spin-up during the end phase of the optically thick mass transfer and the model confirmed that the rejuvenation of the gainer star is principally produced by mass transfer.

\item The hydrogen enrichment on the gainer star causes rejuvenation of the accretor and a relocation within the main sequence. This process causes the gainer to be more massive and compacted that the expected for a star of the same mass.

\item The luminosity of the donor decreases one order of magnitude during 0.18 Myr, due that the deeper layers need expand for adjusting quickly the mass lost, while the gainer star increases the temperature by 10000 K, causing an increase of one order of magnitude of the luminosity, but both processes are not related to the DPV second photometric variation.

\item The central density of the donor increases gradually between mass transfer (D) and the end of the optically thick mass transfer (U$_{2}$).


\item Outside of convective and overshooting mixing zones, the advection has an important role on the generation of currents mechanic, probably causing the generation of seed magnetic field and the appearance of magnetic fields within of donor star. Therefore the convection is not necessary for the generation of the stellar dynamos.

\item The best model, indicate that the DPV V495 Cen is found with age 301.7636 Myr, M$_{d}=0.913$ M$_{\odot}$ and M$_{g}=5.651$ M$_{\odot}$, and is the result of the evolution and mass transfer of two stars of initial masses $M_{i,d}=3.40$ $M_{\odot}$ and $M_{i,g}=3.18$ $M_{\odot}$ and initial orbital period 3.9 days.

{\item Since most DPVs have an early B-type gainer and a A/F/G giant donor, they probably follow similar evolutionary routes starting with a pair of main sequence, intermediate mass stars, orbiting the center of mass in few days, for evolving into a semidetached configuration after the evolution and radius increase of the initially more massive component  (\citet{2016MNRAS.461.1674M}). This work suggests that the case study of V 495 Cen can be observed as representative  for the enter population of  DPVs, or at least, for most of these systems.}

\end{itemize}

\vskip 0.5cm
{\bf ACKNOWLEDGEMENTS}\\

This research was funded in part by a scholarship from Faculty of Physical Sciences and Mathematics and  Dirección de Postgrado of the Universidad de Concepción (UdeC). J.R. and R.E.M gratefully acknowledges support from the Chilean BASAL Centro de Excelencia en Astrofísica y Tecnologías Afines (CATA) grant PFB-06/2007 and AFB-170002.  A.S. acknowledges to Bishop's University for giving the BEST Project Fund, which helped to attend to {\textsc{mesa}} Summer School 2017 and collaborate in this work. R.E.M. acknowledges support by grant VRID 218.016.004-1.0. DRGS thanks for funding via Fondecyt regular (project code 1161247) and through the ''Concurso Proyectos Internacionales de Investigaci\'on, Convocatoria 2015'' (project code PII20150171).



\begin{thebibliography}{99}

\bibitem[Applegate \& Patterson(1987)]{1987ApJ...322L..99A} Applegate, J.~H., \& Patterson, J.\ 1987, apjl, 322, \href{http://dx.doi.org/10.1086/185044}{L99}
\bibitem[Baker \& Kippenhahn(1959)]{1959ZA.....48..140B} Baker, N., \& Kippenhahn, R.\ 1959, zap, 48, \href{http://adsabs.harvard.edu/abs/1959ZA.....48..140B}{140} 
\bibitem[Biermann(1950)]{1950ZNatA...5...65B} Biermann, L.\ 1950, Zeitschrift Naturforschung Teil A, 5, \href{http://adsabs.harvard.edu/abs/1959ZA.....48..140B}{65} 
\bibitem[Bloecker(1995)]{1995A&A...297..727B} Bloecker, T.\ 1995, aap, 297, \href{http://adsabs.harvard.edu/abs/1995A\%26A...297..727B}{727} 
\bibitem[Burger \& Katz(1983)]{1983ApJ...265..393B} Burger, H.~L., \& Katz, J.~I.\ 1983, apj, 265, \href{http://adsabs.harvard.edu/abs/1983ApJ...265..393B}{393} 
\bibitem[Caleo et al.(2016)]{2016MNRAS.460..338C} Caleo, A., Balbus, S.~A., \& Tognelli, E.\ 2016, mnras, 460, \href{http://adsabs.harvard.edu/abs/2016MNRAS.460..338C}{338} 
\bibitem[Charbonnel \& Zahn(2007)]{2007A&A...476L..29C} Charbonnel, C., \& Zahn, J.-P.\ 2007, aap, 476, \href{http://adsabs.harvard.edu/abs/2007A\%26A...476L..29C}{L29} 
\bibitem[de Jager et al.(1988)]{1988A&AS...72..259D} de Jager, C., Nieuwenhuijzen, H., \& van der Hucht, K.~A.\ 1988, aaps, 72, \href{http://adsabs.harvard.edu/abs/1988A\%26AS...72..259D}{259} 
\bibitem[Eggenberger et al.(2010)]{2010A&A...519A.116E} Eggenberger, P., Meynet, G., Maeder, A., et al.\ 2010, aap, 519, \href{http://adsabs.harvard.edu/abs/2010A\%26A...519A.116E}{A116} 
\bibitem[Eggleton(1971)]{1971MNRAS.151..351E} Eggleton, P.~P.\ 1971, mnras, 151, \href{http://adsabs.harvard.edu/abs/1971MNRAS.151..351E}{351} 
\bibitem[Eggleton(2006)]{2006epbm.book.....E} Eggleton, P.\ 2006, Evolutionary Processes in Binary and Multiple Stars, by Peter Eggleton, pp.~.~ISBN 0521855578.~Cambridge, UK: Cambridge University Press,  \href{http://adsabs.harvard.edu/abs/2006epbm.book.....E}{2006}
\bibitem[Fricke(1968)]{1968ZA.....68..317F} Fricke, K.\ 1968, zap, 68, \href{http://adsabs.harvard.edu/abs/1968ZA.....68..317F}{317} 
\bibitem[Garrido et al.(2013)]{2013MNRAS.428.1594G} Garrido, H.~E., Mennickent, R.~E., Djura{\v s}evi{\'c}, G., et al.\ 2013, mnras, 428, \href{http://adsabs.harvard.edu/abs/2013MNRAS.428.1594G}{1594}
\bibitem[Glebbeek et al.(2009)]{2009A&A...497..255G} Glebbeek, E., Gaburov, E., de Mink, S.~E., Pols, O.~R., \& Portegies Zwart, S.~F.\ 2009, aap, 497, \href{http://adsabs.harvard.edu/abs/2009A\%26A...497..255G}{255} 
\bibitem[Goldreich \& Schubert(1967)]{1967ApJ...150..571G} Goldreich, P., \& Schubert, G.\ 1967, apj, 150, \href{http://adsabs.harvard.edu/abs/1967ApJ...150..571G}{571} 
\bibitem[Heger et al.(2000)]{2000ApJ...528..368H} Heger, A., Langer, N., \& Woosley, S.~E.\ 2000, apj, 528, \href{http://adsabs.harvard.edu/abs/2009A\%26A...497..255G}{368} 
\bibitem[Heger et al.(2005)]{2005ApJ...626..350H} Heger, A., Woosley, S.~E., \& Spruit, H.~C.\ 2005, apj, 626, \href{http://adsabs.harvard.edu/abs/2005ApJ...626..350H}{350} 
\bibitem[Kato(1966)]{1966PASJ...18..374K} Kato, S.\ 1966, pasj, 18, \href{http://adsabs.harvard.edu/abs/1966PASJ...18..374K}{374} 
\bibitem[Kippenhahn \& Weigert(1967)]{1967ZA.....65..251K} Kippenhahn, R., \& Weigert, A.\ 1967, zap, 65, \href{http://adsabs.harvard.edu/abs/1967ZA.....65..251K}{251} 
\bibitem[Kippenhahn(1974)]{1974IAUS...66...20K} Kippenhahn, R.\ 1974, Late Stages of Stellar Evolution, 66, \href{http://adsabs.harvard.edu/abs/1974IAUS...66...20K}{20} 
\bibitem[Kippenhahn et al.(2012)]{2012sse..book.....K} Kippenhahn, R., Weigert, A., \& Weiss, A.\ 2012, Stellar Structure and Evolution: , Astronomy and Astrophysics Library.~ISBN 978-3-642-30255-8.~Springer-Verlag Berlin Heidelberg, \href{http://adsabs.harvard.edu/abs/2012sse..book.....K}{2012}
\bibitem[Kolb \& Ritter(1990)]{1990A&A...236..385K} Kolb, U., \& Ritter, H.\ 1990, A\&A, 236, \href{http://adsabs.harvard.edu/abs/1990A\%26A...236..385K}{385} 
\bibitem[Lauterborn(1970)]{1970A&A.....7..150L} Lauterborn, D.\ 1970, aap, 7, \href{http://adsabs.harvard.edu/abs/1970A\%26A.....7..150L}{150} 
\bibitem[Leal-Ferreira et al.(2013)]{2013A&A...554A.134L} Leal-Ferreira, M.~L., Vlemmings, W.~H.~T., Kemball, A., \& Amiri, N.\ 2013, aap, 554, \href{http://adsabs.harvard.edu/abs/2013A\%26A...554A.134L}{A134} 
\bibitem[Lebreton et al.(2009)]{2009CoAst.158..277L} Lebreton, Y., Montalb{\'a}n, J., Godart, M., et al.\ 2009, Communications in Asteroseismology, 158, \href{http://adsabs.harvard.edu/abs/2009CoAst.158..277L}{277} 
\bibitem[Maeder(2009)]{2009pfer.book.....M} Maeder, A.\ 2009, Physics, Formation and Evolution of Rotating Stars: , Astronomy and Astrophysics Library.~ISBN 978-3-540-76948-4.~Springer Berlin Heidelberg, \href{http://adsabs.harvard.edu/abs/2009pfer.book.....M}{2009}
\bibitem[Meakin \& Arnett(2007)]{2007ApJ...667..448M} Meakin, C.~A., \& Arnett, D.\ 2007, apj, 667, \href{http://adsabs.harvard.edu/abs/2007ApJ...667..448M}{448} 
\bibitem[Meintjes(2004)]{2004MNRAS.352..416M} Meintjes, P.~J.\ 2004, mnras, 352, \href{http://adsabs.harvard.edu/abs/2004MNRAS.352..416M}{416} 
\bibitem[Mennickent(2017)]{2017SerAJ.194....1M} Mennickent, R.~E.\ 2017, Serbian Astronomical Journal, 194, \href{http://adsabs.harvard.edu/abs/2017SerAJ.194....1M}{1} 
\bibitem[Mennickent et al.(2016)]{2016MNRAS.461.1674M} Mennickent, R.~E., Zharikov, S., Cabezas, M., \& Djura{\v s}evi{\'c}, G.\ 2016, mnras, 461, \href{http://adsabs.harvard.edu/abs/2016MNRAS.461.1674M}{1674} 
\bibitem[Mennickent et al.(2012)]{2012IAUS..282..317M} Mennickent, R.~E., Ko{\l}aczkowski, Z., Djura{\v s}evi{\'c}, G., Diaz, M., \& Niemczura, E.\ 2012, From Interacting Binaries to Exoplanets: Essential Modeling Tools, 282, \href{http://adsabs.harvard.edu/abs/2012IAUS..282..317M}{317} 
\bibitem[Mennickent et al.(2003)]{2003A&A...399L..47M} Mennickent, R.~E., Pietrzy{\'n}ski, G., Diaz, M., \& Gieren, W.\ 2003, A\&A, 399, \href{http://adsabs.harvard.edu/abs/2003A\%26A...399L..47M}{L47} 
\bibitem[Nieuwenhuijzen \& de Jager(1990)]{1990A&A...231..134N} Nieuwenhuijzen, H., \& de Jager, C.\ 1990, aap, 231, \href{http://adsabs.harvard.edu/abs/2003A\%26A...399L..47M}{134} 
\bibitem[Nugis \& Lamers(2000)]{2000A&A...360..227N} Nugis, T., \& Lamers, H.~J.~G.~L.~M.\ 2000, aap, 360, \href{http://adsabs.harvard.edu/abs/2000A\%26A...360..227N}{227} 
\bibitem[Paczynski \& Proszynski(1986)]{1986ApJ...302..519P} Paczynski, B., \& Proszynski, M.\ 1986, apj, 302, \href{http://adsabs.harvard.edu/abs/1986ApJ...302..519P}{519} 
\bibitem[Paxton(2004)]{2004PASP..116..699P} Paxton, B.\ 2004, pasp, 116, \href{http://adsabs.harvard.edu/abs/2004PASP..116..699P}{699} 
\bibitem[Paxton et al.(2011)]{2011ApJS..192....3P} Paxton, B., Bildsten, L., Dotter, A., et al.\ 2011, apjs, 192, \href{http://adsabs.harvard.edu/abs/2011ApJS..192....3P}{3} 
\bibitem[Paxton et al.(2013)]{2013ApJS..208....4P} Paxton, B., Cantiello, M., Arras, P., et al.\ 2013, apjs, 208, \href{http://adsabs.harvard.edu/abs/2013ApJS..208....4P}{4} 
\bibitem[Paxton et al.(2015)]{2015ApJS..220...15P} Paxton, B., Marchant, P., Schwab, J., et al.\ 2015, apjs, 220, \href{http://adsabs.harvard.edu/abs/2015ApJS..220...15P}{15} 
\bibitem[Paxton et al.(2018)]{2018ApJS..234...34P} Paxton, B., Schwab, J., Bauer, E.~B., et al.\ 2018, apjs, 234, \href{http://adsabs.harvard.edu/abs/2018ApJS..234...34P}{34} 
\bibitem[Peterson et al.(2010)]{2010Natur.463..207P} Peterson, W.~M., Mutel, R.~L., G{\"u}del, M., \& Goss, W.~M.\ 2010, nat, 463, \href{http://adsabs.harvard.edu/abs/2010Natur.463..207P}{207} 
\bibitem[Prialnik \& Kovetz(1995)]{1995ApJ...445..789P} Prialnik, D., \& Kovetz, A.\ 1995, apj, 445, \href{http://adsabs.harvard.edu/abs/1995ApJ...445..789P}{789} 
\bibitem[Regev et al.(2016)]{2016mfdp.book.....R} Regev, O., Umurhan, O.~M., \& Yecko, P.~A.\ \href{http://adsabs.harvard.edu/abs/2016mfdp.book.....R}{2016}, Modern Fluid Dynamics for Physics and Astrophysics.~Series: Graduate Texts in Physics, ISBN: ISBN 978-1-4939-3163-7/ISBN.~Springer New York (New York, NY), Edited by Oded Regev, Orkan M.~Umurhan and Philip A.~Yecko,  
\bibitem[Reimers(1975)]{1975MSRSL...8..369R} Reimers, D.\ 1975, Memoires of the Societe Royale des Sciences de Liege, 8, \href{http://adsabs.harvard.edu/abs/1975MSRSL...8..369R}{369} 
\bibitem[Rosales Guzm{\'a}n et al.(2018)]{2018MNRAS.476.3039R} Rosales Guzm{\'a}n, J.~A., Mennickent, R.~E., Djura{\v s}evi{\'c}, G., Araya, I., \& Cur{\'e}, M.\ 2018, mnras, 476, \href{http://adsabs.harvard.edu/abs/2018MNRAS.476.3039R}{3039} 
\bibitem[Sarna et al.(1997)]{1997MNRAS.286..209S} Sarna, M.~J., Muslimov, A., \& Yerli, S.~K.\ 1997, mnras, 286, \href{http://adsabs.harvard.edu/abs/1997MNRAS.286..209S}{209} 
\bibitem[Schleicher \& Mennickent(2017)]{2017A&A...602A.109S} Schleicher, D.~R.~G., \& Mennickent, R.~E.\ 2017, A\&A, 602, \href{http://adsabs.harvard.edu/abs/2017A\%26A...602A.109S}{A109}
\bibitem[Soberman et al.(1997)]{1997A&A...327..620S} Soberman, G.~E., Phinney, E.~S., \& van den Heuvel, E.~P.~J.\ 1997, aap, 327, \href{http://adsabs.harvard.edu/abs/1997A\%26A...327..620S}{620} 
\bibitem[Soderhjelm(1980)]{1980A&A....89..100S} Soderhjelm, S.\ 1980, aap, 89, \href{http://adsabs.harvard.edu/abs/1980A\%26A....89..100S}{100} 
\bibitem[Spruit(2002)]{2002A&A...381..923S} Spruit, H.~C.\ 2002, aap, 381, \href{http://adsabs.harvard.edu/abs/2002A\%26A...381..923S}{923} 
\bibitem[Tassoul(2000)]{2000stro.book.....T} Tassoul, J.-L.\ 2000, Stellar rotation / Jean-Louis Tassoul.~Cambridge ; New York : Cambridge University Press, \href{http://adsabs.harvard.edu/abs/2000stro.book.....T}{2000}.~(Cambridge astrophysics series ; 36)  
\bibitem[Tassoul(1978)]{1978trs..book.....T} Tassoul, J.-L.\ 1978, Princeton Series in Astrophysics, Princeton: University Press, \href{http://adsabs.harvard.edu/abs/1978trs..book.....T}{1978}  
\bibitem[Ulrich(1972)]{1972ApJ...172..165U} Ulrich, R.~K.\ 1972, apj, 172, \href{http://adsabs.harvard.edu/abs/1972ApJ...172..165U}{165} 
\bibitem[van Rensbergen et al.(2008)]{2008A&A...487.1129V} van Rensbergen, W., De Greve, J.~P., De Loore, C., \& Mennekens, N.\ 2008, A\&A, 487, \href{http://adsabs.harvard.edu/abs/2008A\%26A...487.1129V}{1129} 
\bibitem[van Rensbergen et al.(2011)]{2011A&A...528A..16V} van Rensbergen, W., de Greve, J.~P., Mennekens, N., Jansen, K., \& de Loore, C.\ 2011, aap, 528, \href{http://adsabs.harvard.edu/abs/2011A\%26A...528A..16V}{A16} 
\bibitem[Vauclair(2008)]{2008IAUS..252...97V} Vauclair, S.\ 2008, The Art of Modeling Stars in the 21st Century, 252, \href{http://adsabs.harvard.edu/abs/2008IAUS..252...97V}{97} 
\bibitem[Vlemmings et al.(2005)]{2005A&A...434.1029V} Vlemmings, W.~H.~T., van Langevelde, H.~J., \& Diamond, P.~J.\ 2005, aap, 434, \href{http://adsabs.harvard.edu/abs/2005A\%26A...434.1029V}{1029} 
\bibitem[Vink et al.(2001)]{2001A&A...369..574V} Vink, J.~S., de Koter, A., \& Lamers, H.~J.~G.~L.~M.\ 2001, aap, 369, \href{http://adsabs.harvard.edu/abs/2001A\%26A...369..574V}{574} 
\bibitem[Wasiutynski(1946)]{1946ApNr....4....1W} Wasiutynski, J.\ 1946, Astrophysica Norvegica, 4, \href{http://adsabs.harvard.edu/abs/1946ApNr....4....1W}{1} 
\bibitem[Zahn(1992)]{1992A&A...265..115Z} Zahn, J.-P.\ 1992, aap, 265, \href{http://adsabs.harvard.edu/abs/1992A\%26A...265..115Z}{115} 

\end{thebibliography}
\end{document}